# Revenue diversification in emerging market banks: implications for financial performance

Saoussen Ben Gamra[*]         Dominique Plihon[**]

**Abstract.**

Shaped by structural forces of change, banking in emerging markets has recently experienced a decline in its traditional activities, leading banks to diversify into new business strategies. This paper examines whether the observed shift into non-interest based activities improves financial performance. Using a sample of 714 banks across 14 East-Asian and Latin-American countries over the post 1997-crisis changing structure, we find that diversification gains are more than offset by the cost of increased exposure to the non-interest income, specifically by the trading income volatility. But this diversification performance's effect is found to be no linear with risk, and significantly not uniform among banks and across business lines. An implication of these findings is that banking institutions can reap diversification benefits as long as they well-studied it depending on their specific characteristics, competences and risk levels, and as they choose the right niche.

[*] CEPN, High Institute of Management of Sousse, Tunisia. Email: saoussengamra@yahoo.fr
[**] CEPN, University of Paris North, France. Email: dplihon@orange.fr

# 1. Introduction

The 1997 financial crisis in emerging market economies has important implications for the feasibility of different banking models. The crisis has clearly exposed the dangers of a bank's excessive reliance on the traditional business activities in a context of capital account liberalization. Many analysts have particularly advanced the lack of proper diversification of the loan portfolio as a key catalyst of bank distress after financial deregulation (Stone, 2000; Radelet and Sachs, 1999). Over time, the structure of emerging countries banking markets has been shaped by policies that encourage the provision of financial services to specific sectors of economies on the fringe of economic development. After the crisis, the universal banking model, which allows banks to combine a wide range of financial activities, including commercial banking, investment banking and insurance, has emerged as a desirable structure for a financial institution from the viewpoint of policymakers. While most banking systems not surprisingly still rely mainly on income from traditional banking, the post-crisis years have seen an increasing number of banks especially in East-Asia and Latin-America moving into investment banking-type activities, fee-based business and related activities (Laeven, 2007). The shifts have been leading to a blurring of lines across different financial institutions and have been facilitated by relatively liberal laws as regards banking and securities business.

Conventional industry wisdom predicts that combining different types of activities –non-interest earning and interest-earning assets– and rebalancing bank income away from interest income and toward non-interest activities may increase return and diversify risks, therefore boosting performance. Empiric's studies tend, however, to contradict these conventional industry beliefs. Research provides conflicting conclusions about a bank's benefits of optimal asset or activity mix. In general, these studies provide only little evidence of diversification benefits. In an early survey, Saunders and Walters (1994) review 18 studies that examine whether non-bank activities reduce bank holding company risk and indicate no consensus: 9 answer yes, 6 answer no, and 3 are mixed.

Few earlier studies find some potential for gain from expansion into specific activities. Boyd et al. (1980), Kwast (1989), Templeton and Severiens (1992), and Gallo et al. (1996) examined combination of US bank and non-bank activities, and find a slight potential for risk reduction at relatively low levels of non-bank activities. Conversely, significant strand of literature related to the performance and diversification of US banks draw a general conclusion that bank expansion into less traditional financial activities is associated with increased risk and/or lower returns. Demsetz and Strahan (1997) show that better diversification does not translate into reductions in overall risk. Kwan (1998) finds that Section 20 subsidiaries are typically more risky but not necessarily more profitable than their commercial bank affiliates. DeYoung and Roland (2001) look at the impact of fee-based activities for 472 large US commercial banks, and conclude to an increase in the volatility of bank revenue and the existence of a risk premium associated with these activities.

Studies that followed generated similar findings. Stiroh (2002, 2006) shows that greater reliance on non-interest income has been associated with higher volatility of bank income and higher risk, but not higher returns. Hirtle and Stiroh (2007) find that the increased focus of US banks on retail banking was associated with significantly lower equity and accounting returns for all banks. This result is consistent with evidence provided by the several studies of DeYoung and Rice (2004a,b,c). Increases in non-interest income at US banking companies are associated with higher, but more volatile, rates of return, resulting in reduced risk-adjusted returns. More relevant recent research (Stiroh and Rumble, 2006; Goddard et al. 2007) based even on US data suggests that expansion into non-traditional activities is associated with more volatile revenue streams that can offset the risk-spreading benefits of diversification. It concludes that the direct exposure effect reflecting the impact of non-interest income dominates an indirect exposure effect related to diversification.

A less uniform pessimistic picture emerges from studies for countries other than the US. A study of loan portfolio diversity by Acharya et al. (2006) reports that diversification of bank assets is not guaranteed to produce superior performance and/or greater safety for banks for a sample of 105 Italian banks. An empirical analysis of Mercieca et al. (2007) focused on a sample of 755 small banks for 15 European countries find no direct diversification benefits within or across business lines, but an inverse association between non-interest income and bank performance. From 15 European Union countries, Smith et al. (2003) find that non-interest income is less stable than interest income for a sample of banks. Based on a set of 734 European banks, Lepetit et al. (2007) show that banks expanding into non-interest income activities, presented higher risk and higher insolvency risk than banks which mainly supplied loans. Results based on large international sample of banks are more conflicting. Laeven and Levine (2007) examined the effects of diversification on the market value of large banks from 42 countries, and find that the market values of diversified banks were lower than those of their specialized counterparts. Demirguç-Kunt and Huizinga (2009) investigates the implications of banks activity and short-term funding strategies for bank risk and returns using a sample of 1334 banks in 101 countries. They conclude that universal banking can be beneficial in terms of diversifying risks and increasing returns, but banking strategies that rely predominantly on generating non-interest income or attracting non-deposit funding are very risky.

Most of the studies on the benefits of banks engaging in a broader scope of activities have focused on the large, complex banking institutions that dominate US banking and relatively little is known about potential diversification benefits for banks in emerging market economies. This is an important issue because these banks play a critical role in the stability of the global financial system. Should emerging countries banks be diversified or focused on a few related fields? Does diversification indeed lead to enhanced performance and, therefore, greater safety for these banks, as traditional portfolio and banking theory would suggest? This paper proposes to investigate this relationship between bank performance and revenue diversification in the changing structure of the emerging economies banking industry during the post-crisis decade. Based on a broad set of 714 banks across 14 East-Asian and Latin-American countries for the period 1997- end 2007, our study documents the dramatic increase in non-interest income at emerging economies banks during the last decade that reflects diversification of banks into non-traditional activities, and examines the implications of these changes for the financial performance of banks as measured by their risk-adjusted outcomes.

The remainder of this paper is organized as follows. Section 2 provides an overview of the main changes shaping the banking industry in the emerging market economies. Section 3 reviews the theoretical literature on diversification, and develops our hypothesis. In section 4, we focus on stylized facts of two different Latin American and Asian economies –Argentina and Korea– to examine the recent evolving to diversification in emerging market banks. Section 5 specifies the variables and describes the data. We present the empirical results in section 6, and offer concluding remarks in section 7.

## 2. The changes in emerging market banking industry

Banking systems in emerging economies have undergone substantial changes in the last few decades. During recent period, banking industry faces a shrinking of the portion of intermediated funds handled by banks and a decline in traditional intermediation activities, leading bank institutions to move into new business strategies as investment banking-type and related fee-generating activities. At least four forces underline this bank shift into non-traditional activities: domestic deregulation, technology innovations, changes in corporate behavior and banking crises.

*2.1. Deregulation and opening-up to foreign competition*

Banking in the emerging economies was traditionally a highly protected industry, living off good spreads achieved on regulated deposit and lending rates and pervasive restrictions on domestic and foreign entry. Global market, technology developments and macroeconomic pressures have forced the banking industry and the regulators to change the old way of doing business, and to deregulate the banking industry at the national level and open up financial markets to foreign competition. As a result, borders between financial products, banks and non-bank financial institutions and the geographical locations of financial institutions have started to break down. These changes have significantly increased competitive pressures on banks in the emerging economies and have led to deep changes in the banking strategies.

One of the main catalysts for increased competition at the domestic level has been the removal of ceilings on deposit rates and the lifting of prohibitions on interest payments on current accounts. These deregulation measures have reduced sources of cheap funding for many banks and put pressure on their traditional intermediation profits. Banks reacted to this margins compression and market power decline by raising their involvement in new activities and diversifying their activities which considerably altered their income structure by reducing the importance of their traditional lines of business. Banks also increasingly face competition from the non-bank financial institutions, especially for lending to large companies. They are thus constrained to expand their product lines and to offer financial products and services that had previously been reserved for other financial institutions. In addition, savers place now a larger share of their savings in other financial institutions like mutual funds and pension funds. Since, many banks cannot acquire all the core deposits they want, they engage in liability management by borrowing in the money market. At the margin, this change in bank liability structure could affect its allocation of resources between traditional and non-traditional activities by producing a larger quantity of non-traditional activities concurrently with finding other sources of funds[1]. Intensified competition has also made it harder for banks to cross-subsidize different activities and has forced them to price risks more realistically and to charge explicitly for previously free services. Accompanying deregulation has been greater emphasis on capital adequacy, which has encouraged banks to securitize some assets and generate more fee-based income.

*2.2. Technology innovation*

New information technology is not at present likely to impinge much on the development of the banking industry in the emerging economies, which remain technologically behind the industrial countries[2]. However, banks are required to exploit the new technologies, which can fundamentally change banking business models. The major issue about new information technology is its impact on the processing of information, which is the very essence of the banking business, and the emergence of entirely new financial instruments and production processes. Banks are required to innovate in services and products, especially new deposit and loan-based offerings, to differentiate strategies and to combine different activities to set themselves apart from rivals. They would need to fundamentally transform its business into a much wider array of non-traditional services. The rising importance of off-balance sheet activities, ranging from credit lines to derivatives products, is symptomatic of this development. In this new technological environment, loans to small and moderate-sized businesses based on private, information-rich relationships between business people and their commercial bankers stand out as one of the last types of loans that are still produced in the traditional intermediation fashion.

---

[1] Unlike, traditional loans, certain non-traditional activities allow banks to provide services without having to obtain balance sheet funding.
[2] For example, the low level of penetration in most emerging economies means that the internet is not seen as a threat to traditional banks. Given the signs of a possible bursting of the e-banking bubble in the United States and Europe, some have also argued that the issue of electronic banking may go away before the emerging markets need to worry about it.

*2.3. Changes in corporate behavior*

The spread of information technology has affected the banking industry both directly, through information technology applications in marketing of financial products, and indirectly, through its impact on the development of financial markets and corporate behavior. This impact is most clearly felt in the case of technology firms, which are more or less forced to turn to capital markets to finance their projects because banks are not prepared to deal with the high level of uncertainty associated with the development of new technologies. However, disintermediation is not limited to new economy firms; it is also beginning to be felt in the more traditional old economy sectors. Larger firms have been moving away from commercial bank loans toward open market securities like commercial paper or long-term bonds. Bonds outstanding have risen strongly in almost all emerging economies over the past few years, allowing many large firms to raise funds by issuing securities more cheaply than they can borrow from banks. Banks are under increasing pressure to keep their customers, and to the extent that more and more creditworthy firms turn to alternative funding sources and the proportion of higher-risk bank customers increases, banks, especially in the emerging economies, are forced to develop techniques for better pricing and provisioning of credit risks. The increased demand for such risk management services meant that in addition to their traditional intermediary role, banks were called upon to provide such services[3]. This meant that banks had to increasingly diversify out of their traditional banking operations and provide fee-based services. There were standard contracts for hedging of risks associated with markets. However, when corporations desired products tailored to their specific needs, they turned to banks for those products. This demand led to a wide variety of custom-tailored contracts such as loan commitments, forward contracts and swaps. The growth of off-balance sheet activities was a natural outgrowth of banks providing such risk management services. In addition, banks have an incentive to increase their presence and role on financial markets by providing both lending and other services for the firm, such as underwriting, guarantees, holding equity and engaging in venture capital activities. This is further encouraged by the development of financial instruments inducing more investment in real assets. Alternatively, the shift to trading-based services could continue, and banks could become more involved as asset gatherers and active intermediaries in these markets.

---

[3] For example, companies that borrowed in their domestic currency, derived income in other currencies from their foreign operations and banks could help such companies to control their foreign currency risk. Similarly, technology-intensive firms for whom unpredictable short-term revenues imposed severe constraints on their research and development (R&D) budgets, approached banks that provided products designed to hedge overseas income and plan R&D over longer period.

*2.4. Systemic banking crises in the 1990s*

There were many systemic banking crises in emerging markets during the 1990s, often occurring shortly after the financial deregulation. Considerable attention in the financial crisis literature has been devoted to macroeconomic and institutional causes of banking crises. In particular, unsustainably high growth of lending to the private sector, poor prudential regulations and bank supervision, and premature capital account liberalization were identified as major contributing factors. However, some of the most common sources of banking crises are microeconomic in nature, including the insufficiently diversified loan books that made specialist banks over-dependent on the particular region or sector served; excessive optimism about lending to rapidly expanding manufacturing firms and speculative property developers; credit assessment by banks often very poor, and loans often made to related companies or state-owned enterprises, frequently at the behest of governments; management incentives often inappropriate; and the risks from excessive maturity and currency mismatches not fully appreciated. The proportion of loans that have become impaired during banking crises in the emerging markets has generally been much greater than that in the industrial world, implying also higher economic costs, especially in the relatively large bank-based financial systems in Asia.

The post-crisis decade has been marked by a typical change in the bank behavior of emerging economies. First, there is often a substantial decline in private sector intermediation that reflects a "flight to liquidity". Banks have restructured their portfolio towards highly liquid public securities, cash reserves and disproportionately decrease private sector credit reflecting the strategy chosen to minimize risk after systemic distress. Second, there is a decline in bank profitability often linked to the high level of non-performing loans on banks' balance sheets. Banks typically get rid of their loans, and find new business lines such as fee-based activities and investment in government securities[4]. From the viewpoint of the regulators and supervisors, the emergence of the universal banking model, which allows banks to combine a wide range of financial activities, is a desirable structure for bank institutions, implicitly assumed to be part of the stabilization process. Non-traditional activities are viewed as helping to reduce the risk of bankruptcy since they will be diversifying the income generated by the bank, which could have a positive effect on firm value.

This evolution of emerging economies' banking underlines the changing nature and structure of the banking industry. As banks attempt to compete in the broader and evolving financial-services industry, they alter their behavior by changing the menu of products and services they offer. These changes can be viewed as part of an overall strategy to expand beyond traditional sources of revenues, to rely less on scarce core deposits as a source of financing, and to help boosting profits and reducing risk exposures. This diversification into non-traditional activities resulting from the evolving business of banking rises questions about its potential implications on bank performance what motivates this paper. It is also interesting from a policy standpoint to investigate this issue.

---

[4] For instance, in Brazil recovery of bank profitability was not a result of greater intermediation *per se,* but of the reorientation of banks portfolios towards liquidity, predominantly government securities.

## 3. Theoretical background and hypothesis

Theory provides conflicting predictions about the impact of greater diversification of activities on the performance of financial intermediaries. Existing theories of financial intermediation imply increasing returns to scale linked to diversification. As suggested by the work of Diamond (1991), Rajan (1992), Saunders and Walter (1994), and Stein (2002), banks acquire customer information during the process of making loans that can facilitate the efficient provision of other financial services, including the underwriting of securities. Similarly, securities and insurance underwriting, brokerage and mutual fund services, and other activities can produce information that improves loan making. Thus, banks that engage in a variety of activities could enjoy economies of scope that boost performance.

There is also a cost linked to intermediary risk, and a better diversified intermediary has less risk and thus lower costs. In models of insurance or liquidity provision (Diamond and Dybvig, 1983; Chari and Jagannathan, 1988; Jacklin and Bhattarcharya, 1988; Gorton and Pennacchi, 1990), investors are risk averse and face some risk which the intermediary can pool and diversify on their behalf. Moreover, diversification makes it cheaper for financial institutions to achieve credibility in their role as screeners or monitors of borrowers. As shown by the models of delegated investment monitoring or evaluation (Campbell and Kracaw, 1980; Diamond, 1984; Ramakrishnan and Thakor, 1984; Boyd and Prescott, 1986), the possibility of bad outcomes allows the intermediary to hide proceeds or to claim that bad luck rather than lack of effort led to the bad outcomes; an intermediary with better diversified investments has less chance of very bad outcomes, reducing associated costs. Thus, that it is optimal for a bank to be maximally diversified across sectors.

Experts of diversification argue also that lenders such as banks and finance companies are typically highly levered, and diversification across sectors reduces their chance of costly financial distress. Similarly, the conventional view is that greater competition has increased the need for banks to diversify: lower profits leave less margin for error, so diversification provides a necessary reduction in risk. Only a simple policy prescription for regulators is suggested by the traditional theory: the banking sector should be left relatively unrestricted, which should in turn lead to an equilibrium with a few large, well-diversified, and competitive banks.

The Winton's models (1997, 1999) results of the proverbial wisdom of "not putting all your eggs on one basket" suggest that the opposite can be true. Increased competition may magnify the "Winner's Curse" problem (the adverse selection in the borrowers pooling) faced on entry into a new sector, making diversification very costly. In unregulated settings where intermediaries are new or the market is growing rapidly, there should be substantial entry, with many risky intermediaries coexisting: investors cannot coordinate their actions, and debt overhang makes the cost of capturing market share through rate competition highest when the potential for diversification is greatest. Over time, banks will fail, and survivors will gain an incumbency advantage simply by becoming the focus of investor beliefs. Banks facing greater competition may therefore find it more attractive to specialize. In related work, several models (Dell'Arricia, Friedman, and Marquez, 1999; Marquez, 1997; Dell'Arricia, 1998; Gehrig, 1998) suggest that regardless of the bank's efforts, loans in the new sector are likely to perform worse than loans in the bank's home sector. Worse performance for new sector loans also makes diversification more likely to increase the bank's chance of failure and less likely to improve the bank's monitoring incentives; indeed, diversification may even undermine incentives to monitor home sector loans. Overall, diversification is more likely to be unattractive.

Considerable literature exists on banks' non-traditional activities, it looks at different financial activities separately and shows that these activities affect differently the level of risk at an individual bank (e.g. Avery and Berger, 1991; Boot and Thakor, 1991; Hassan, 1992, 1993; Hassan et al. 1994; Hassan and Sackley, 1994). By definition, diversification involves moving into economic sectors that differ from the bank's home base. Effective loan monitoring requires that the lending institution have a thorough understanding of these differences, but building such organizational knowledge takes time and effort. Alternatively, diversification of activities within a single financial conglomerate could intensify agency problems between corporate insiders and small shareholders (Jensen, 1986; Jensen and Meckling, 1976). Since, it is difficult for outsiders to directly observe the lending process that a bank is following, with adverse implications on the market valuation of the conglomerate.

Hence, we formulate the first hypothesis: **H1.** *A bank's monitoring effectiveness may be lower in newly entered and competitive sectors, and thus, diversification can result in a poorer quality of loans that in turn increase the bank's loan portfolio risk and reduce return.*

There has been some work on bank specialization and loan performance. A somewhat closer study is Besanko and Thakor (1993), who model insured banks allocating loans across two uncorrelated sectors. Diversified banks forfeit gains from risk-shifting but increase their odds of surviving to collect informational rents on continuing lending relationships; free entry reduces these rents, discouraging diversification. In addition to the winner's curse problem facing new entrants, Boot and Thakor (1998) examine incentives to specialize in the face of increased competition.

Nonfinancial corporate diversification literature (Denis et al., 1997; Rajan, Servaes and Zingales, 2000; Maksimovic and Philips, 2002) generally argues that any firm –financial institution or other– should focus on a single line of business so as to take greatest advantage of management's expertise and reduce agency problems, leaving investors to diversify on their own (Jensen, 1986; Berger and Ofek, 1996; Servaes, 1996, Denis et al., 1997). Linked corporate literature regarding the "diversification discount" finds also that the market value of financial conglomerates that engage in multiple activities are lower than if those financial conglomerates where broken into financial intermediaries that specialize in the individual activities. According to Demsetz and Strahan (1997), the diversification discount may be caused by that too many operating items make the banks lose their focus on specialized field. Another reason may cause the diversification discount including the inefficient internal resource allocation (Lamont, 1997; Scharfstein, 1997), the informational asymmetries between head office and divisional managers (Harris, Kriebel and Raviv, 1992).

But the features that distinguish banks and other lenders from nonfinancial firms are lenders' greater use of debt finance (leverage) and the way in which lenders' efforts affect their return distributions. With high leverage, worst-case outcomes loom large both in terms of underinvestment problems and in terms of outright failure. Although pure diversification tends to reduce the frequency of both worst-case and best-case outcomes, diversification that lessens monitoring effectiveness may increase the frequency and severity of worst-case outcomes, increasing failure probability and underinvestment problems (Winton, 1999). Furthermore, Winton (1999) consider that "pure" diversification increases the central tendency of the bank's return distribution, which generally reduces the bank's chance of failure. Nevertheless, if its loans have sufficiently low exposure to sector downturns ("downside"), a specialized bank has a low probability of failure, so the benefit of diversification is slight. Also, if its loans have sufficiently high downside, diversification can actually increase the bank's chance of failure. Thus, all else equal, diversification's benefits are greatest when the bank's loans have moderate levels of downside risk and when the bank's monitoring incentives need strengthening.

We formulate the second hypothesis: **H2.** *The relationship between bank return and diversification is non linear in bank risk (inverted U-shaped). To be precise, diversification across loan sectors helps a banks return most when loans have moderate exposure to sector downturns (downside risk); when loans have low downside risk, diversification have little benefit; when loans have sufficiently high downside risk, diversification may actually reduce its return.*

Broadly speaking, diversification per se is no guarantee of a reduced risk of failure and/or an increased return. Contrasting views suggest that neither diversification nor specialization always dominates; some circumstances and bank specific differences can favor one strategy or the other. More generally, "diversification discount" models predict that firms can differ in terms of expansion opportunities capabilities and ability to exploit market occasions. For example, the Maksimovic and Phillips model (2002) of optimal resource allocation of firms shows that as a firm's returns within an industry diminish, the firm limits its growth within the industry and moves into other industries. The optimal number and size of industry segments a firm operates depends on its comparative advantage across industries, arising from managerial skill in producing within an industry. Firms that are very productive in a specific industry have higher opportunity costs of diversifying. Thus, inefficient and efficient firms should optimally invest differently when industry conditions change.

Similarly, greater size is required for better diversification at the same time large institutions have substantial scale economies linked to improved diversification (Roger and Sinkey, 1999). Participation in certain non-traditional activities generally requires employees with special knowledge to work in some of these areas. Moreover, a bank might need to employ relatively advanced technology for some activities. Larger banks are better equipped to use new technology and exploit the resulting cost savings and/or efficiency gains (Hunter and Timme, 1986).

A more diversified bank may have also greater relative need for equity capital, especially if diversification involves expansion into sectors where the bank is less effective (Winton, 1995). Banks do use debt for much of their financing, equity capital serves as a buffer to absorb losses and reduce the probability of financial distress. In addition, by reducing possible shortfalls on payments to debtholders, equity capital reduces the bank's incentive to engage in risk-shifting by not monitoring. Also, high bank profits can be seen as to reduce the likelihood of costly bank runs and bank default resulting from bank involvement into new activities.

In another way, Barth et al. (2004) arguments' for restricting activities suggest that it improves the banking system by avoiding banks from the problems like conflicts of interest, complexity, moral hazard and monitoring difficulties. As banks expand to new activities, the restrictions may direct banks to less risky and less complicated activities and thus improve bank diversification performance. However, if this is not the case, the restrictions may misdirect banks to riskier and more complicated activities and thus decrease diversification performance.

Lastly we formulate the third hypothesis: **H3.** *The diversification performance's effect is inherently different by activities and across banks. There are some situations where financial institutions gains greatly from diversification, but this depends on diverse bank specific characteristics, as well as regulatory measures.*

# 4. Evolving to diversification in emerging market banks: the stylized facts from Argentina and Korea

In this section, we examine the evolution that marked recently emerging markets' banking systems. We focus on stylized facts of two different Latin American and Asian economies – Argentina and Korea–, which faced banks evolving to diversification of their strategies and revenues, with a significant shift toward activities that generate non-interest income.

*3.1. Banking diversification and performance in Argentina*

*3.1.1. Diversification strategies*

The Argentine convertibility plan ended in January 2002 in the middle of a huge social, financial and economic crisis. With a new set of relative prices in place, the economy started its recovery in the second half of 2002. The positive evolution of the economy then allowed the financial sector to begin its recovery from the crisis, after a period of bank intermediation contraction. The demand for credit fell because of recession and the greater reluctance of borrowers to become indebted. At the same time, the supply of credit declined: banks become more risk-averse and a major stiffening of supervisory oversight reinforced this effect.

After the crisis, bank credit begun to rise again slowly. Loans grow to 46 percent of the total assets against 40 percent of the other-earning assets at the median bank at the end of 2007 (Tab.1). Precisely, the share of credit going to the business sector has declined sharply from 48 percent to 25 percent of total loans between the late of 1997 and 2007. The demand for commercial credit declined. Highly rated firms are increasingly able to borrow directly in domestic and international capital markets, reducing their reliance on banks. Banks may also ultimately respond to this corporate diversification by moving into consumer lending and investment banking-type activities. The favorable prospects for economic growth for 2005 and 2006 have set up an encouraging scenario for the consumer lending. Consumer loans showed a robust growth from 1 percent to 18 percent of total loans over the period. However, banks engaged more in securities investment. Its share grew considerably to about 25 percent of the other-earning assets at the end of 2007, while deposits with banks grew slowly and other investments declined. Off-balance sheet items also increased. This evolution reflects the banks moving into investment-banking and related activities like securities underwriting and trading, securitization, and derivatives.

**Tab.1. Argentina: Elements of the banking system's balance sheet**

|  | 31/12/1997 | 31/12/2007 |
|---|---|---|
| *As % of total assets* | | |
|     Loans | 45.01 | 46.24 |
|     Other earning assets | 45.04 | 40.70 |
|     Deposit & ST funding | 86.44 | 77.68 |
|     Other funding | 0.30 | 1.88 |
|     Equity | 11.81 | 13.71 |
|     Off-balance sheet items | 0.76 | 10.66 |
| *As % of total loans* | | |
|     Commercial loans & bills | 48.37 | 25.00 |
|     Consumer loans | 1.16 | 18.84 |
|     Secured loans | 12.70 | 7.59 |
| *As % of total other earning assets* | | |
|     Deposits with banks | 4.45 | 9.86 |
|     Securities | 7.83 | 24.94 |
|     Other investments | 32.17 | 19.62 |

*Notes*: Median value percentages for a sample of 85 Argentine banks.    *Source*: Authors' calculation from Bankscope.

This shift toward market related activities has been leading to a blurring of lines across different financial institutions and has been facilitated by liberal laws as regards banking and securities business. Securities expanded as a cause of the substantial accumulation of holdings of government or central bank securities. Increased issuance of government securities has many counterparts: larger fiscal deficits, bank recapitalization and the increased local currency financing of budget deficits. The market for innovative financial products has continued to expand, and banks have increased their off-shore positions over the past decade. Still low interest rates fueled the growth of core deposits but the rise was insufficient to fund the increase in banks assets. As a result, banks relied more on managed liabilities, which begun to rise slightly last year –most notably long-term borrowing.

### 3.1.2. Bank income and performance

The income structure of the banking system provides a clear breakdown of the activities and income of Argentine banks by lines of business. The extent to which the banking system has been moving into investment-banking type business is shown by the diversification of their income (Tab.2). While most banks in the country not surprisingly still rely on income from traditional banking, the importance of non-traditional business income has increased and is relatively high. Realized gains on non-interest activities boost income by 50 percent at median in all banks at the end of 2007.

**Tab.2. Argentina: Structure of the banking system's income**

|  | 31/12/1997 | 31/12/2000 | 31/12/2007 |
|---|---|---|---|
| *As % of operating income* | | | |
| Net interest revenue | 60.13 | 56.52 | 46.97 |
| Non-interest income | 39.86 | 43.47 | 50.02 |
| *As % of non-interest income* | | | |
| Net fees and commissions | 80.90 | 44.25 | 53.04 |
| Other non-interest income | 19.09 | 55.74 | 47.07 |
| *including* | | | |
| Government & private securities | 12.79 | | 10.83 |
| Foreign exchange Transactions | 6.68 | | 2.51 |
| Other | 0.00 | | 81.94 |

*Notes*: Median value percentages for a sample of 85 Argentine banks.   *Source*: Authors' calculation from Bankscope.

Importantly, gains are greater in retail banks and reached 53 percent of total income at the end-2007. As a share of total revenue, interest income sharply decreases from 60 percent to about 47 percent between the late of 1997 and 2007 at the median bank. The share of fees and commissions income falls, in favor of the other non-interest revenue share expanded. Loans to the financial sector, promissory notes, mortgages and credit cards grew especially fast throughout the period, when cash income has contracted to a very low level. Banks increased the share of these loans in their portfolios. Investing in these assets, which appear to have relatively higher yields than commercial loans, allowed them to limit the decline in the overall rate of return on their assets.

Tab.3 shows that banking system is very profitable at the end of 2007. Pre-tax profit to assets ratio, return on assets, and return on equity were all high and strongly above the level prevailing before the 2001 crisis. Profits become from the non-interest activities because the net interest margin declined a bit further at these banks from 1 percent between the late of 1997 and 2007. Banks rely on managed liabilities for their funding. Because rates paid on these liabilities are more sensitive to changes in market interest rates than are rates paid on core deposits, the net interest margin was adversely affected. The net interest margin was also eroded by continued intensification of competitive pressure and runoffs of their commercial loans. The share of interest-earning assets attributable to such loans fell. The large drop was only partially offset by a shift toward higher-yielding loans, such as credit card loans. Banks also increased their share of interest-earning assets that consisted of investment-account securities. Because, rates of return on securities are generally lower than those on loans, this shift contributed further to the narrowing of the net interest margin.

Competition could lower financial intermediation costs and contribute to improvements in bank efficiency as shown by the cost to income ratio. The bank efficiency in non-interest services has also improved but costs relied to these services remained higher relatively to those of the other traditional services. Equity to assets and equity to loans ratios surged at the end-2007, and remains above the pre-crisis period level, which means that banks have riskier assets. New non-traditional activities in which banks engaged are more generators of profits than traditional activities, but more risky. The continued improvement in overall credit quality driven by the strengthening of household and business balance sheets and the ongoing expansion, has allowed banks to further reduce their provisions for loans losses.

**Tab.3. Argentina: Banking system's financial ratios (%)**

|  | 31/12/1997 | 31/12/2000 | 31/12/2003 | 31/12/2007 |
|---|---|---|---|---|
| Pre-tax profit to total assets | 0.9 | 0.9 | -0.9 | 2.3 |
| Return on assets (ROA) | 1.22 | 0.42 | -1.33 | 1.92 |
| Return on equity (ROE) | 10.31 | 3.38 | -5.02 | 13.35 |
| Net interest margin | 6.04 | 5.83 | 3.01 | 5.07 |
| Cost to income | 75.39 | 76.63 | 94.6 | 71.06 |
| Non-interest expenses/ Total revenue | 80.60 | 84.07 | 115.87 | 72.69 |
| Capital funds to total assets | 10.55 | 12.53 | 21.81 | 11.88 |
| Capital funds to net loans | 22.64 | 28.12 | 60.18 | 23.53 |
| Equity to assets | 11.4 | 13.3 | 20.75 | 14.3 |
| Equity to loans | 26.2 | 29.9 | 65.6 | 31.4 |
| Loan loss provisions to loans | 2.31 | 2.91 | 0.92 | 0.62 |

*Notes*: Median value percentages for a sample of 85 Argentine banks.  *Source*: Authors' calculation from Bankscope.

*3.2. Banking diversification and performance in Korea*

*3.2.1. Diversification strategies*

The Korean banking industry has undergone substantial structural reforms since the 1997 financial crisis. A number of banks merged or exited and foreign banks were permitted to enter the banking industry. Healthy banks, meanwhile, were strongly encouraged to seek consolidation and develop universal banking services in order to become leaders in the banking industry. The biggest and most significant merger took place in November 2001, when Kookmin Bank & Commercial Bank merged to create Korea's largest commercial bank. Because of their strengths and market positions, the merger of the two banks has generated a new wave of competitive pressures and contributed to further restructuring and consolidation in the banking sector.

Recapitalization of the financial sector was needed to relieve a paralyzing credit crunch at the beginning of the crisis and to restore the proper functioning of banks in their role as financial intermediaries. Loans expanded at median bank to 62 percent of total assets at the end of 2007, with a doubling of currency domestic lending share to about 80 percent of total loans (Tab.4). More interesting, investment and traded securities represent a large part of the other-earning assets of Korean banks. Off-balance sheet items also increased at the end of 2007 compared to assets (27 percent). The saving and time deposits fall sharply, when other sources of funding expanded and become almost exclusively formed by bonds and notes at 98 percent last year. Balance sheet data point to relatively large holding of securities in relation to total assets. This evolution reflects a greater market orientation of Korean banks regarding their activities and revenues as well as their financing. Banks tend to diversify their strategies by moving into new capital-market activities and combining them with the traditional intermediation functions. Two factors underline this shift. One is the development of the capital markets and the increasing availability of direct financing. The other is the change in the behavior and risk exposure of banks to corporate. This led them increasingly to emphasize an expansion of their market-focused activities.

**Tab.4. Korea: Elements of the banking system's balance sheet**

|  | 31/12/1997 | 31/12/2007 |
|---|---|---|
| *As % of total assets* | | |
| Loans | 51.03 | 62.08 |
| Other earning assets | 38.26 | 26.65 |
| Deposit & ST funding | 71.07 | 71.24 |
| Other funding | 7.36 | 10.45 |
| Equity | 7.56 | 7.73 |
| Off-balance sheet items | 18.69 | 27.18 |
| *As % of total loans* | | |
| Notes & bills discounted | 6.53 | 0.94 |
| Currency domestic lending | 44.47 | 79.84 |
| Foreign currency & overseas lending | 6.70 | 2.78 |
| Other lending (incl. Leasing) | 8.40 | 7.26 |
| *As % of total other earning assets* | | |
| Deposits with banks | 15.60 | 20.12 |
| Traded securities | 0.86 | 4.53 |
| Investment securities | 46.40 | 51.90 |
| Equity investments | 2.27 | 0.93 |

*Notes*: Median value percentages for a sample of 35 Korean banks.     *Source*: Authors' calculation from Bankscope.

### 3.2.2. Bank income and performance

Diversification in bank strategies into new market activities reflects a major ongoing shift in the structure of the Korean banking income. While, the interest revenue remains dominant in this structure, its part is in declining in favor of the non-interest income that rose to almost 40 percent of the total operating income at the median bank in end-2007 (Tab.5). Loans and discounts contributed strongly to the interest income, with a share of 78 percent at the end of 2007, but interest-income from securities increased last year to 13 percent.

The structure of the non-interest income has also shifted toward trading income. The net fees and commissions contracted last year to 52 percent against 63 percent in late 1997, when the net trading share increases significantly to 27 percent at the median bank. Important gains are realized on investment securities, trading securities, and derivatives, respectively 32, 9 and 8.4 percent of the trading income at the end-2007. Indeed, the function of banks' fund distribution to the national economy has changed radically. Banks avoid cut-throat competition in specific asset markets and generate new lucrative activities by expanding their investment banking and finance business.

**Tab.5. Korea: Structure of the banking system's income**

|  | 31/12/1997 | 31/12/2000 | 31/12/2007 |
|---|---|---|---|
| *As % of operating income* |  |  |  |
| Net interest revenue | 63.15 | 62.50 | 60.30 |
| Non-interest income | 36.84 | 38.07 | 39.69 |
| *As % of non-interest income* |  |  |  |
| Net fees and commissions | 63.02 | 66.90 | 52.84 |
| Net trading | 8.90 | 13.24 | 27.36 |
| *including* |  |  |  |
| Net gain on traded securities | 0.00 |  | 9.48 |
| Net profit on foreign ex. Transactions | 100.00 |  | 11.30 |
| Net gain on investment securities | 0.00 |  | 32.63 |
| Net profit on derivatives | 0.00 |  | 8.46 |
| Other non-interest income | 57.40 | 26.64 | 7.62 |

*Notes*: Median value percentages for a sample of 35 Korean banks. *Source*: Authors' calculation from Bankscope.

Strategically, banks need to heighten their profitability by maintaining appropriate interest rate spreads that incorporate borrowers' credit risks and performing better market activities. Banks continued to pile up large deficits after the crisis until 2000 because they suffered from the losses incurred by non-performing loans, but they have shown profits since 2001. Strikingly, banks reached a pre-provisions profit to total assets of 1.80 percent, and had an average ROA of 1.17 percent and ROE of 14.37 percent in late 2007 (Tab.6). One of the most crucial elements of the improvement of profitability is that it has been possible to reduce new provisions against bad loans (0.35 percent in end-2007).

**Tab.6. Korea: Banking system's financial ratios (%)**

|  | 31/12/1997 | 31/12/2000 | 31/12/2003 | 31/12/2007 |
|---|---|---|---|---|
| Pre-provisions profit to total assets | 0.99 | 1.15 | 1.50 | 1.80 |
| Return on assets (ROA) | 0.29 | 0.13 | 0.40 | 1.17 |
| Return on equity (ROE) | 4.38 | 4.03 | 3.71 | 14.37 |
| Net interest margin | 1.95 | 2.07 | 2.49 | 2.47 |
| Cost to income | 75.20 | 61.88 | 59.12 | 58.49 |
| Non-interest expenses/ Total revenue | 75.20 | 55.25 | 47.25 | 51.00 |
| Capital funds to total assets | 4.98 | 6.38 | 7.35 | 8.76 |
| Capital funds to net loans | 8.32 | 12.01 | 11.22 | 13.93 |
| Equity to assets | 7.60 | 4.20 | 4.80 | 7.70 |
| Equity to loans | 9.49 | 7.64 | 8.19 | 12.94 |
| Loan loss provisions to loans | 2.92 | 2.80 | 1.65 | 0.35 |

*Notes*: Median value percentages for a sample of 35 Korean banks. *Source*: Authors' calculation from Bankscope.

The build-up of interest bearing assets and the increase of non-interest income from trading, contributed to improving banks' profitability, in spite of the narrowing net interest margin compared to the emerging Asia average. Bank efficiency, capitalization and leverage have significantly improved since the financial crisis, but capital ratios are still high and comparables to those of the crisis level. The level of equity to assets and equity to loans ratios of Korean banks reveals also high risk of the held assets. This trend is driven by the increase of banks' new market activities involvement. Korean banks appear relatively efficient in these activities as shown by the ratio of non-interest expenses to total revenue that has fallen considerably last year. But, investing in business, which appears to have relatively larger and rapid yields than traditional bank intermediation activities, allowed them to take and expose themselves to more risk.

## 5. Variables, data and statistics

### 5.1. Variables definition

#### 5.1.1. Diversification measures

To measure the income diversification, we follow the basic Herfindhal-type approach. Our measure of revenue diversification (DIV$_{REV}$), accounts for variation in the breakdown of net operating income into two broad categories: net interest income and non-interest income. Using this breakdown, we measure revenue diversification of the emerging economies banks as:

$DIV_{REV} = 1 - (SH_{NET}^2 + SH_{NON}^2)$. Where SH$_{NET}$ is the share of net interest income, and SH$_{NON}$ is the share of non-interest income, defined as: $SH_{NET} = \frac{NET}{NET+NON}$; $SH_{NON} = \frac{NON}{NET+NON}$.

The non-interest income includes commissions and fees income (FEE), and trading and other income (TRAD), which the respective shares are defined as: $SH_{TRAD} = \frac{TRAD}{NON}$; $SH_{FEE} = \frac{FEE}{NON}$.

In an analogous fashion, we defined three measures of revenue diversification forms: assets diversification as $DIV_{ASS} = 1 - (SH_{LOANS}^2 + SH_{OEA}^2)$, where SH$_{LOANS}$ is the share of loans, and SH$_{OEA}$ is the share of other earning assets (non-lending activities); liabilities or funding diversification as $DIV_{FUND} = 1 - (SH_{DEPOSITS}^2 + SH_{OF}^2)$, where SH$_{DEPOSITS}$ is the share of deposits, and SH$_{OF}$ is the share of other money market funding; and balance sheet diversification as $DIV_{BAL} = 1 - (SH_{ONBAL}^2 + SH_{OFBAL}^2)$, where SH$_{ONBAL}$ is the share of on-balance sheet, and SH$_{OFBAL}$ is the share of off-balance sheet.

Diversification variables measure the degree of bank diversification. A higher value indicates a more diversified mix. The value 0 means a complete concentration, while 0.5 means a complete diversification.

*5.1.2. Financial performance measures*

We use three types of risk-adjusted performance measures: risk-adjusted return on assets, risk-adjusted return on equity or Sharpe ratio, and Z-score. Annual accounting data from banks balance sheet are utilized to calculate these indicators for each bank individually. Risk-adjusted returns are defined as the ratio of return divided by its respective standard deviation as follows: $RAR_{ROA} = \overline{ROA}/\sigma_{ROA}$ and $RAR_{ROE} = \overline{ROE}/\sigma_{ROE}$. Higher ratios indicate higher risk-adjusted profits. Z-score assesses insolvency risk as follows: $Z = (\overline{ROA} + \overline{E/A})/\sigma_{ROA}$. Where $\overline{E/A}$ is the average of equity to assets ratio. Thus, a higher Z-score indicates improved risk-adjusted performance and lower probability of bank insolvency. It's interpreted as the distance to default or the number of standard deviation that a bank's rate of return of assets has to fall for the bank to become insolvent.

*5.1.3. Control variables*

We use three categories of control variables: the operational environment variables, bank specific variables, and other dummy variables. Operational environment variables include the concentration indicator that measures the competition faced by banks (low index indicates greater competition); and the bank freedom index that measures how much latitude a bank has to make operating decisions. It is an indicator of relative openness of banking and financial system. Bank specific variables control for bank characteristics and differences in the structure and strategy that can be expected to affect a bank's income mix as well as risk and return outcomes. First, the log of total assets (Log TA) is used to proxy for bank size and to control for any systematic differences in performance across size classes, e.g., scale economies, or different risk-management techniques. Second, equity to assets ratio (E/A) is included to measure bank capitalization, and the risk preferences of banks, i.e., risk loving banks may hold less equity. Third, the interest share captures the percentage of traditional activities. Finally, dummy variables are included for each country, each bank type and for the number of years the bank is observed.

## 5.2. Data and descriptive statistics

All bank-level data in this study are taken from Bankscope database, which is currently the most comprehensive banks data set. The sample period is from 31/12/1997 to 31/12/2007. Banks with less than five years of time series observations are eliminated. Other observations with extreme or non-sensical values are deleted. In total, 714 banks established in 14 Latin-America and East-Asia countries are surveyed (Appendix. Table1). Our overall sample includes banks of different sizes and types, although 73% of the sample is comprised of large commercial bank observations (Appendix.Table2).

*5.2.1. Trends on non-interest income share, and bank risk-adjusted performance*

Figure1 shows the growing importance of non-interest income for emerging economies banks over time by plotting the non-interest income share. Banks have clearly shifted toward non-interest activities. From 1997 to late 2007, banks saw non-interest income as a share of net operating revenue rise from 28.2% to stabilize around 36.7% (an increase of 30%). While not surprisingly, small banks and very small banks saw an increase of 47% and 57% respectively (Appendix.Fig.1A). Steep increases in the non-interest share by end-2007 are seen mainly for investment banks and securities houses (Appendix.Fig.1B). The key point is that banks of all sizes and types are shifting toward non-interest income, and not just limited to a few mega-banks that perform many diverse activities. The biggest increase of non-interest income was in trading and other revenue that grew the fastest from 30% of non-interest income in 1997 to 47% by end-2007. This shows also the changing focus of bank strategies within non-interest income.

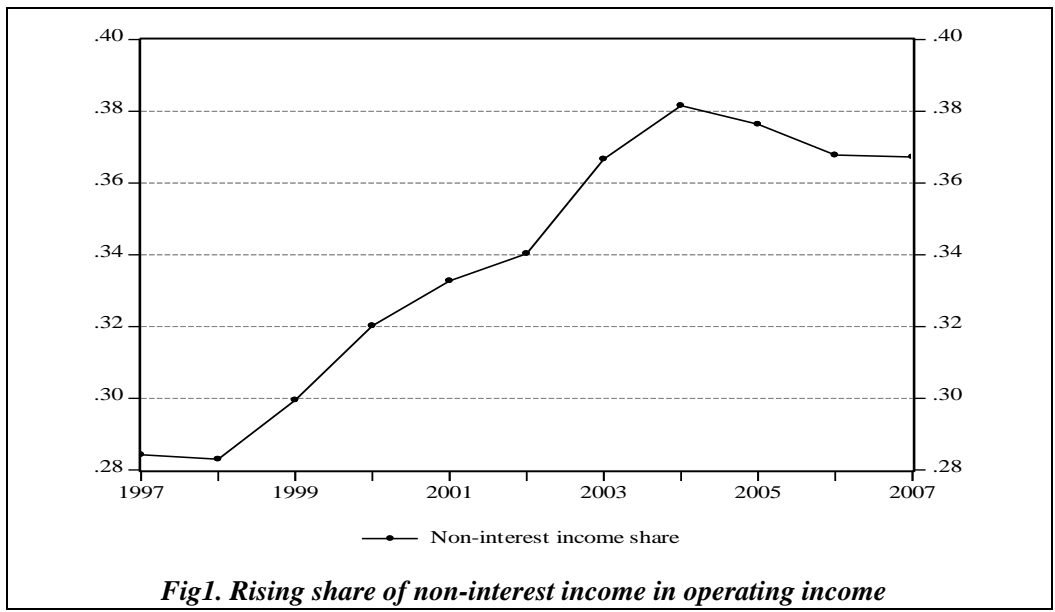

*Fig1. Rising share of non-interest income in operating income*

Figure 2 provides a graphical representation of the relationship between the bank risk-adjusted performance as measured by the Sharpe ratio and the Z-score, and the non-interest income. In both cases, figure show a strong negative slope across the non-interest income bins. The risk-adjusted measures declined steadily. Banks with high non-interest income shares seem to earn lower risk-adjusted profits and are relatively risky.

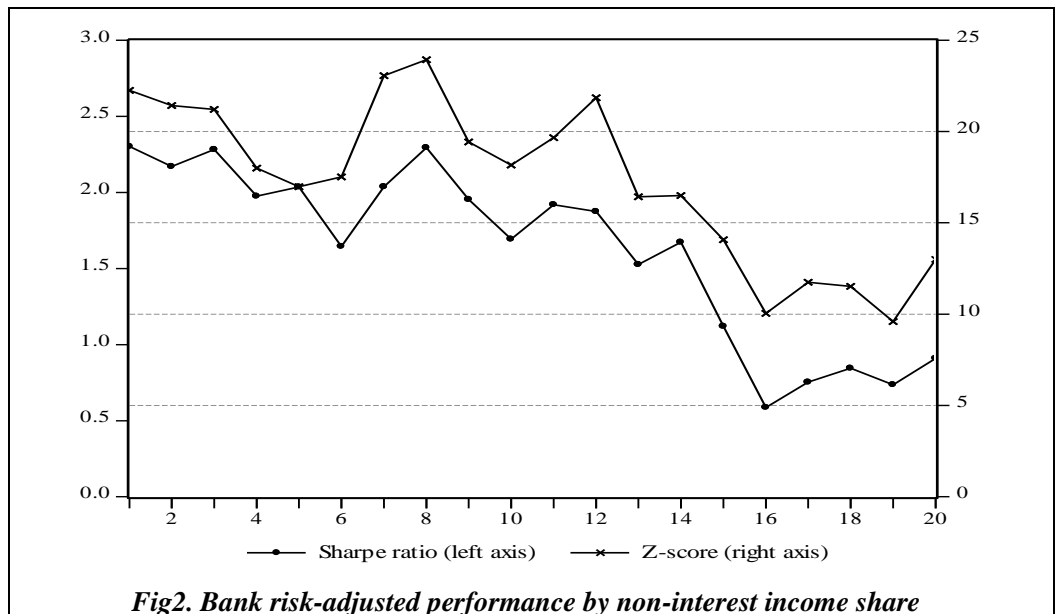

*Fig2. Bank risk-adjusted performance by non-interest income share*

Notes: We reported risk-adjusted performance as measured only by the Sharpe ratio and the Z-score because graphs on risk-adjusted return on assets and equity are almost similar. Both measures are averaged for all banks in each bin, where bins are created by sorting banks by their average non-interest income share and making 20 equal sized groups containing 5 percent of observations of these variables in increasing order.

### 5.2.2. *Bank-level summary statistics*

Table7 displays summary statistics of the main variables used for the 714 emerging countries banks in this sample. The banks were quite varied and averaged about $6.77 billion in assets. This sample is dominated by a few large banks so the mean greatly exceeds the median. Diversification revenue measure ranges from total concentration in net-interest income (0.00) to total diversification (0.50). But, the distribution of average diversification for the main sample of 714 emerging market banks shows an asymmetric distribution with a fatter right-hand tail. This indicates not surprisingly that the majority of banks are somewhat diversified in terms of net operating revenue. Relatively few banks are seen to rely almost exclusively on non-interest income or net-interest income. In fact, the distribution of the non-interest income share peaks for values of this variable between 0.40 and 0.45, the overall sample mean of the non-interest income share is 0.34 (Appendix.Fig.2). Comparing different forms of revenue diversification indicates that emerging market banks are more diversified in terms of assets (0.35) than in terms of funding (0.09) or balance sheet (0.14). The share of non-earning assets remains significant (46.5% on average), and banks focused recently more equally on trading activities, and commissions and fees. On the performance side, our sample includes both low- and high-performance banks. The mean risk-adjusted return on assets and equity was 1.59 and 1.65, with a range from -1.21 to 13.17 and from -1.48 to 14.8 respectively; while the mean Z-score was 17.28, with a range from 0.73 to 122.84. Table 8 presents correlation matrix to check for possible collinearity among the variables. All variables included in the model are weakly correlated with each.

**Table7. Summary statistics for emerging economies banks**

|  | Obs | Mean | Median | Std. Dev | Minimum | Maximum |
|---|---|---|---|---|---|---|
| *Bank characteristics* | | | | | | |
| Total Assets ($ millions) | 714 | 6777.7 | 838.36 | 0.40 | 7.4 | 284841.4 |
| Equity to Assets (%) | 714 | 20.28 | 13.21 | 0.17 | 2.77 | 97.48 |
| Interest share (%) | 714 | 66.00 | 70.00 | 0.20 | 0.00 | 100.00 |
| *Operational environment* | | | | | | |
| Concentration (%) | 714 | 62.52 | 60.80 | 0.25 | 19.50 | 100.00 |
| Bank freedom | 714 | 53.71 | 49.09 | 14.59 | 34.54 | 90.00 |
| *Diversification revenue* | | | | | | |
| $DIV_{Rev}$ | 714 | 0.33 | 0.35 | 0.11 | 0.00 | 0.50 |
| $DIV_{Assets}$ | 714 | 0.35 | 0.40 | 0.12 | 0.00 | 0.49 |
| $DIV_{Fund}$ | 714 | 0.09 | 0.04 | 0.11 | 0.00 | 0.49 |
| $DIV_{Bal}$ | 714 | 0.14 | 0.11 | 0.13 | 0.00 | 0.49 |
| *Non-traditional shares* | | | | | | |
| $SH_{Non}$ (%) | 714 | 34.00 | 30.00 | 0.20 | 0.00 | 97.00 |
| $SH_{OEA}$ (%) | 714 | 46.55 | 44.39 | 0.22 | 0.00 | 100.00 |
| $SH_{OF}$ (%) | 714 | 8.88 | 2.67 | 0.15 | 0.00 | 100.00 |
| $SH_{OFBAL}$ (%) | 714 | 12.75 | 6.88 | 0.16 | 0.00 | 94.10 |
| *Risk-adjusted performance* | | | | | | |
| $RAR_{ROA}$ (%) | 714 | 1.59 | 1.16 | 0.01 | -1.21 | 13.17 |
| Sharpe ratio ($RAR_{ROE}$) (%) | 714 | 1.65 | 1.24 | 0.01 | -1.48 | 14.80 |
| Z-score (%) | 714 | 17.28 | 11.70 | 0.16 | 0.73 | 122.84 |

**Table8. Correlation matrix between main variables**

|  | LOG TA | E/A | Concentration | Bank freedom | $DIV_{REV}$ | $SH_{NON}$ | $RAR_{ROA}$ | Sharpe ratio | Z-score |
|---|---|---|---|---|---|---|---|---|---|
| LOG TA | 1 | | | | | | | | |
| E/A | -0.554 | 1 | | | | | | | |
| Concentration | -0.124 | -0.010 | 1 | | | | | | |
| Bank freedom | 0.031 | 0.065 | 0.496 | 1 | | | | | |
| $DIV_{REV}$ | 0.174 | -0.211 | 0.171 | 0.051 | 1 | | | | |
| $SH_{NON}$ | -0.071 | 0.174 | 0.068 | 0.113 | 0.366 | 1 | | | |
| $RAR_{ROA}$ | 0.310 | -0.161 | 0.085 | 0.011 | -0.057 | -0.263 | 1 | | |
| Sharpe ratio | 0.304 | -0.170 | 0.101 | -0.009 | -0.052 | -0.267 | 0.878 | 1 | |
| Z-score | 0.132 | 0.055 | 0.160 | 0.102 | -0.033 | -0.202 | 0.747 | 0.609 | 1 |

Our statistical analysis shows a considerable variation in bank strategy and outcomes. This helps to identify differences in operating outcomes and performance by differences in the diversification measures. To be more formal about the relationship between financial performance and bank reliance on non-interest income, we may utilize appropriate empirical regressions.

## 6. Empirical results

We investigate empirically our three theoretical hypotheses exposed in section 3. Different equations are estimated with OLS. This is essential because all variables are calculated over time for each bank as a combination of means and standard deviations for all years the bank is observed.

### *6.1. The effect of revenue diversification on bank performance*

We test empirically our first theoretical hypothesis **H1**: *A bank's monitoring effectiveness may be lower in newly entered and competitive sectors, and thus, diversification can result in a poorer quality of loans that in turn increase the bank's loan portfolio risk and lower return*. We explore this issue by examining first how the overall variation in bank strategic focus affects their performance, and then we assess more specifically the impact of the different components of the non-interest income as well as the different forms of revenue diversification on bank performance.

#### *6.1.1. Revenue diversification, non-interest income and bank performance*

The first regressions examine the overall effect of revenue diversification and non-interest income on bank performance. Results are presented in Table9. The control variables coefficients appear largely reasonable. Greater bank concentration that means lower competition is associated with better performance reported to risk, while financial freedom that allows banks and other financial institutions to enter and to operate freely, reduce bank profitability. Size is positively associated with return adjusted to risk, which is consistent with the recent scale economies literature (Hughes et al., 2001). The positive coefficient on the capital ratio likely signals the risk-aversion of banks; relatively safer banks have both high capital ratios and low volatility of income and risk.

Results on diversification variables show a strong negative correlation between revenue diversification and bank risk-adjusted return, which means that banks which exhibit high degrees of diversification into non-traditional income display lower return and higher profits' volatility and risk. The coefficient of the non-interest income share itself is negative and highly significant in all regressions, suggesting that the negative effect of the revenue diversification on return and risk are driven by the greater exposure to the non-interest income share.

Table 9. The effect of revenue diversification and non-interest income on risk-adjusted performance

| Regressors | Risk-adjusted return | | Z-score |
|---|---|---|---|
| | $RAR_{ROA}$ | Sharpe ratio | |
| **$DIV_{REV}$** | -0.0153*** | -0.0150*** | -0.0208 |
| | [0.0055] | [0.0056] | [0.0551] |
| **$SH_{NON}$** | -0.0196*** | -0.0197*** | -0.1844*** |
| | [0.0024] | [0.0025] | [0.0223] |
| Bank concentration | 0.0113*** | 0.0138*** | 0.0970*** |
| | [0.0037] | [0.0033] | [0.0295] |
| Bank freedom | -0.0001** | -0.0001*** | 0.0001 |
| | [0.0000] | [0.0000] | [0.0004] |
| Equity to assets | 0.0095*** | 0.0086*** | 0.2335*** |
| | [0.0034] | [0.0036] | [0.0427] |
| Log total assets | 0.0073*** | 0.0072*** | 0.0433*** |
| | [0.0009] | [0.0011] | [0.0078] |
| Constant | -0.0111*** | -0.0086** | -0.1375*** |
| | [0.0045] | [0.0045] | [0.0406] |
| Observations (Adj $R^2$) | 714 (0.20) | 714 (0.20) | 714 (0.16) |

Notes: Regressions used Y = [$RAR_{ROA}$, Sharpe ratio, Z-score] as dependant variables. The estimated equation is:
$Y_i = \alpha + \beta DIV_{REV_i} + \delta SH_{NON_i} + \gamma X_i + \varepsilon_i$. Where $SH_{NON}$ is the share of non-interest income; $DIV_{REV}$ is the revenue diversification's variable; X is the vector of control variables; and ε is the error terms. Equation is estimated with OLS. Dummy variables for country, years and bank type are included in all regressions but not reported. ***, ** indicate significance at 99% and 95% level of significance respectively.

Globally, estimates indicate that substituting 1% of non-interest revenue for loan revenue would reduce significantly ROA adjusted to risk by about 1.67% at the median bank, and reduce Sharpe ratio significantly by about 1.59%, while insolvency risk raise significantly by about 1.57%[5]. These findings support recent work of DeYoung and Roland (2001), and Stiroh and Rumble (2006) who show that diversification gains are more than offset by the cost of increased exposure to the non-interest activities volatility. These activities may ultimately be profitable, but adjustment costs could hold down the short run returns, that they called "the dark side of the search to diversify". Banks in emerging markets may need time to build the business practices, scale, technology, and expertise to successfully combine these different products sometimes risky and achieve high-risk adjusted profits.

### 6.1.2. Components of non-interest income and bank performance

The impact of diversification within non-interest income is split into two components, the trading and other income effect, and the commissions and fees income effect. Results of regressions reported in Table 10 show that the bank exposure to these two business lines is significantly different. Fee income increase does not necessarily affect bank outcomes, inversely it can generate some improvement in the Sharpe ratio, while high reliance on trading income imply lower performance as measured by risk-adjusted returns. Trading income is the most volatile part of non-interest income (Stiroh, 2006). Hence, it is reasonable that the negative effect of non-interest income on bank return/risk seems essentially driven by the trading income share. This trading result is consistent with Estrella (2001), who finds that merger between banking and securities firms are less likely to produce gains because of the securities firms' highly volatile returns.

---
[5] Using $SH_{NON}$ estimates in Table 9, we report in percentage the coefficient found in each regression multiplied by 1% to the corresponding median risk-adjusted return.

Table10. The effect of different components of non-interest income on risk-adjusted performance

| Regressors | Risk-adjusted return | | | | Z-score | |
|---|---|---|---|---|---|---|
| | $RAR_{ROA}$ | | Sharpe ratio | | | |
| **$SH_{NON}$** | **-0.0249***** | **-0.0261***** | **-0.0218***** | **-0.0239***** | **-0.2657***** | **-0.2630***** |
| | [0.0033] | [0.0035] | [0.0030] | [0.0033] | [0.0292] | [0.0300] |
| **$SH_{TRAD}$** | **-0.0072***** | | **-0.0098***** | | **-0.0117** | |
| | [0.0030] | | [0.0029] | | [0.0264] | |
| **$SH_{FEE}$** | | **0.0024** | | **0.0054*** | | **-0.0237** |
| | | [0.0034] | | [0.0033] | | [0.0284] |
| Bank concentration | 0.0145*** | 0.0130*** | 0.0170*** | 0.0149*** | 0.1146*** | 0.1144*** |
| | [0.0042] | [0.0042] | [0.0038] | [0.0036] | [0.0336] | [0.0345] |
| Bank freedom | -0.0000* | -0.0000 | -0.0001*** | -0.0001** | 0.0001 | 0.0001 |
| | [0.0000] | [0.0000] | [0.0000] | [0.0000] | [0.0005] | [0.0005] |
| Equity to assets | 0.0061 | 0.0036 | 0.0076* | 0.0042 | 0.1618*** | 0.1576*** |
| | [0.0044] | [0.0043] | [0.0042] | [0.0042] | [0.0396] | [0.038 3] |
| Log total assets | 0.0058*** | 0.0051*** | 0.0062*** | 0.0053*** | 0.0268*** | 0.0260*** |
| | [0.0011] | [0.0011] | [0.0011] | [0.0010] | [0.0090] | [0.0085] |
| Observations (Adj $R^2$) | 714 (0.21) | 714 (0.20) | 714 (0.24) | 714 (0.23) | 714 (0.22) | 714 (0.22) |

Notes: Regressions used Y = [$RAR_{ROA}$, Sharpe ratio, Z-score] as dependant variables. Two equations are estimated:
(1) $Y_i = \propto + \beta SH_{NON_i} + \delta SH_{TRAD_i} + \gamma X_i + \varepsilon_i$ ; (2) $Y_i = \propto + \beta SH_{NON_i} + \delta SH_{FEE_i} + \gamma X_i + \varepsilon_i$
Where $SH_{NON}$ is the share of non-interest income; $SH_{TRAD}$ is the share of trading income; $SH_{FEE}$ is the share of fee income; X is the vector of control variables; and ε is the error terms. Equations are estimated with OLS. Dummy variables for country, years and bank type are included in all regressions but not reported. ***, **, * indicate significance at 99%, 95% and 90% level of significance respectively.

### *6.1.3. Forms of diversification and bank performance*

The impact of revenue diversification is decomposed into three effects: the assets diversification effect, the liabilities (funding) diversification effect and the balance sheet diversification effect. Results of regressions reported in Table11 indicate that at diverse aspects of diversification are associated different implications for bank risk-adjusted performance.

On the assets diversification's side, the coefficient of diversification is far from statistical significance in all specifications, indicating no impact on adjusted to risk performance from changes in the bank assets diversification. Interestingly, the non-earning assets share remains negative and statistically significant at 99% level of significance in all three specifications. Results indicate that diversification between lending and non-lending activities itself does not drive performance. Potential risk-spreading benefits of diversification seem more than outweighed by the increase in volatility of the non-lending revenue. Contrarily to the revenue from traditional lending activities that is likely to be stable over time, because switching information costs make it costly for either borrowers or lenders to walk away from a lending relationship, revenue from non-traditional activities are relatively more volatile as supported by many studies (DeYoung and Roland, 2001; Stiroh, 2006).

Table11. The effect of assets, funding and balance sheet diversification, and their respective non-traditional income shares on bank risk-adjusted performance

| Regressors | Risk-adjusted return | | Z-score |
|---|---|---|---|
| | RAR$_{ROA}$ | Sharpe ratio | |
| **DIV $_{ASS}$** | **-0.0066** | **0.0052** | **-0.0494** |
| | [0.0049] | [0.0046] | [0.0421] |
| **SH $_{OEA}$** | **-0.0080***** | **-0.0068***** | **-0.0843***** |
| | [0.0026] | [0.0025] | [0.0233] |
| Bank concentration | 0.0079** | 0.0105*** | 0.0714*** |
| | [0.0037] | [0.0034] | [0.0297] |
| Bank freedom | -0.0001*** | -0.0002*** | -0.0007* |
| | [0.0000] | [0.0000] | [0.0004] |
| Equity to assets | 0.0013 | 0.0005 | 0.1526*** |
| | [0.0036] | [0.0037] | [0.0419] |
| Log total assets | 0.0050*** | 0.0049*** | 0.0242*** |
| | [0.0009] | [0.0010] | [0.0080] |
| Interest share | 0.0120*** | 0.0128*** | 0.0879*** |
| | [0.0023] | [0.0023] | [0.0214] |
| Observations (Adj R$^2$) | 714 (0.14) | 714 (0.15) | 714 (0.11) |
| **DIV $_{FUND}$** | **-0.0094** | **-0.0118*** | **-0.0290** |
| | [0.0077] | [0.0072] | [0.0637] |
| **SH $_{OF}$** | **0.0044** | **0.0033** | **0.0085** |
| | [0.0047] | [0.0045] | [0.0434] |
| Bank concentration | 0.0077** | 0.0104*** | 0.0672** |
| | [0.0038] | [0.0034] | [0.0301] |
| Bank freedom | -0.0001*** | -0.0002*** | -0.0006 |
| | [0.0000] | [0.0000] | [0.0004] |
| Equity to assets | 0.0057* | 0.0055* | 0.0919** |
| | [0.0033] | [0.0034] | [0.0405] |
| Log total assets | 0.0040*** | 0.0042*** | 0.0133** |
| | [0.0008] | [0.0009] | [0.0067] |
| Interest share | 0.0129*** | 0.0139*** | 0.0915*** |
| | [0.0023] | [0.0023] | [0.0220] |
| Observations (Adj R$^2$) | 714 (0.13) | 714 (0.15) | 714 (0.09) |
| **DIV $_{BAL}$** | **-0.0188***** | **-0.0227***** | **-0.1572** |
| | [0.0077] | [0.0082] | [0.1626] |
| **SH $_{OFBAL}$** | **0.0159***** | **0.0247***** | **0.0888** |
| | [0.0060] | [0.0070] | [0.1565] |
| Bank concentration | 0.0077** | 0.0103*** | 0.0701*** |
| | [0.0037] | [0.0033] | [0.0297] |
| Bank freedom | -0.0001*** | -0.0002*** | -0.0006 |
| | [0.0000] | [0.0000] | [0.0004] |
| Equity to assets | 0.0054* | 0.0055* | 0.0846*** |
| | [0.0031] | [0.0031] | [0.0349] |
| Log total assets | 0.0041*** | 0.0043*** | 0.0143** |
| | [0.0008] | [0.0009] | [0.0066] |
| Interest share | 0.0119*** | 0.0125*** | 0.0902*** |
| | [0.0023] | [0.0023] | [0.0210] |
| Observations (Adj R$^2$) | 714 (0.14) | 714 (0.16) | 714 (0.10) |

Notes: Regressions used Y = [RAR$_{ROA}$, Sharpe ratio, Z-score] as dependant variables. Three equations are estimated:
(1)  $Y_i = \alpha + \beta DIV_{ASS_i} + \delta SH_{OEA_i} + \gamma X_i + \varepsilon_i$  ;   (2)  $Y_i = \alpha + \beta DIV_{FUND_i} + \delta SH_{OF_i} + \gamma X_i + \varepsilon_i$
(3)  $Y_i = \alpha + \beta DIV_{BAL_i} + \delta SH_{OFBAL_i} + \gamma X_i + \varepsilon_i$. Where DIV$_{ASS}$ is the assets diversification's variable; SH$_{OEA}$ is the share of the other earning assets; DIV$_{FUND}$ is the funding diversification's variable; SH$_{OF}$ is the share of the other money market funding; DIV$_{BAL}$ is the balance sheet diversification's variable; SH$_{OFBAL}$ is the share of the off-balance sheet; X is the vector of control variables; and ε is the error terms. Equations are estimated with OLS. Dummy variables for country, years and bank type are included in all regressions but not reported. ***, **, * indicate significance at 99%, 95% and 90% level of significance respectively.

This real negative volatility impact from non-lending income is combined with a "new toy effect" due to changes in management behavior as management shifts focus toward newly-acquired segments (Schoar, 2002).

On the liabilities diversification's side, estimates show a weak negative correlation between bank funding strategies mix and risk-adjusted performance, which is only found to be statistically significant at 90% level of significance on the Sharpe ratio regression. Results suggest that combining deposit and non-deposit money market funding contributes to alleviate bank leverage and to reduce their profits. Mainly because rates paid on the money-market funding are higher, volatile and more sensitive to changes in market interest rates than are paid on core deposits, the bank return can be adversely affected. The coefficient of the other funding share is nevertheless estimated to have an insignificant impact on the bank risk-adjusted performance. This can be interpreted by the relatively low dependence of emerging market banks on managed liabilities.

The balance sheet diversification's results indicate however that diversification variable enters risk-adjusted return regressions with negative and statistically significant signs. By mixing on and off-balance sheet activities, banks may increase their risk-taking and reduce their ability to better conduct balance sheet activities in favor of other activities quite diversified in nature and beyond their core expertise and knowledge, like securities clearance and brokerage activities, data processing services, and investment and management advisory services (Hassan, 1993). Off-balance sheet items include also loan commitments, financial futures and options contracts, standby letters of credit, foreign exchange contracts, and other derivative products. Although, banks focus on such branches can be highly profitable, as shown by the strong positive coefficient found for the off-balance sheet share. Most obviously, these new activities involve risks which are difficult to quantify and requires certain economies of specialization and know-how.

In sum, our results validate the first hypothesis H1. Findings are consistent with a theory that predicts deterioration in bank monitoring quality and return upon lending expansion into newer or competitive industries. A robust empirical result that emerges from the analysis and corroborates Acharya et al., (2006) is that diversification of income is not guaranteed to produce superior performance and/or greater safety for banks.

## *6.2. The relationship between revenue diversification, risk and bank performance*

We study empirically our second theoretical hypothesis **H2**: *The relationship between bank return and diversification is non-linear in bank risk (inverted U-shaped)*. To examine the nature of this relationship, we conduct estimations on risk-adjusted performance using first the overall revenue diversification's measure, and second the different diversification variables.

First regressions results are reported in Table12. Control variables are generally significant and have the expected signs in all regression specifications. The coefficients on the diversification revenue variable used directly and as a quadratic, are negative and positive respectively, and are statistically significant at 99% level of significance. This holds for all three measures of bank risk-adjusted performance. These results provide support for the U-shape hypothesis (H2) describing the relationship between diversification and performance, conditional on the risk level of the bank. Results are interpreted as diversification has a slight benefit at low bank risk levels, has maximum benefits at moderate risk levels and hurts bank return at very high risk levels.

Table12. The non-linear relationship between revenue diversification and risk-adjusted performance

| Regressors | Risk-adjusted return | | Z-score |
|---|---|---|---|
| | $RAR_{ROA}$ | Sharpe ratio | |
| **$DIV_{REV}$** | **-0.0899***** | **-0.0891***** | **-0.5524***** |
| | [0.0243] | [0.0252] | [0.2351] |
| **$(DIV_{REV})^2$** | **0.1286***** | **0.1278***** | **0.9169***** |
| | [0.0418] | [0.0428] | [0.3808] |
| Bank concentration | 0.0123*** | 0.0148*** | 0.1041*** |
| | [0.0037] | [0.0033] | [0.0295] |
| Bank freedom | -0.0001** | -0.0001*** | 0.0000 |
| | [0.0000] | [0.0000] | [0.0004] |
| Equity to assets | 0.0103*** | 0.0094*** | 0.2390*** |
| | [0.0034] | [0.0036] | [0.0425] |
| Log total assets | 0.0074*** | 0.0073*** | 0.0438*** |
| | [0.0009] | [0.0011] | [0.0078] |
| Interest share | 0.0201*** | 0.0202*** | 0.1879*** |
| | [0.0024] | [0.0026] | [0.0225] |
| Constant | -0.0232*** | -0.0209*** | -0.2684*** |
| | [0.0060] | [0.0062] | [0.0522] |
| Observations (Adj $R^2$) | 714 (0.20) | 714 (0.21) | 714 (0.16) |

Notes: Regressions used Y = [$RAR_{ROA}$, Sharpe ratio, Z-score] as dependant variables. The estimated equation is:
$Y_i = \alpha + \beta DIV_{REV_i} + \delta DIV^2_{REV_i} + \gamma X_i + \varepsilon_i$. Where $DIV_{REV}$ is the revenue diversification variable used directly and as a quadratic; X is the vector of control variables; and ε is the error terms. The quadratic term of $DIV_{REV}$ is incorporated to detect an expected inverted-U shape relationship. Equation is estimated with OLS. Dummy variables for country, years and bank type are included in all regressions but not reported. ***, ** indicate significance at 99% and 95% level of significance respectively.

The second regressions examine the relationship between diversification, risk and bank performance, with considering the different forms of diversification –assets diversification, funding diversification and balance sheet diversification. Earlier results are qualitatively robust to different diversification variables as seen in Table13. The non-linear relationship between diversification, risk and bank performance continues to hold though the coefficients are less significant statistically than under Table12. More generally, assets and balance sheet diversification appear to increase performance at moderate levels of risk, but reduce performance at very high levels of risk.

Table 13. The non-linear relationship between different forms of diversification and bank performance

| Regressors | Risk-adjusted return | | Z-score |
|---|---|---|---|
| | $RAR_{ROA}$ | Sharpe ratio | |
| **$DIV_{ASS}$** | **-0.0636***  | **-0.0439**  | **-0.2739** |
| | [0.0210] | [0.0204] | [0.1725] |
| **$(DIV_{ASS})^2$** | **0.1206*** | **0.0879*** | **0.5801*** |
| | [0.0382] | [0.0366] | [0.3200] |
| Bank concentration | 0.0107*** | 0.0131*** | 0.0962*** |
| | [0.0037] | [0.0034] | [0.0290] |
| Bank freedom | -0.0000* | -0.0001*** | 0.0001 |
| | [0.0000] | [0.0000] | [0.0004] |
| Equity to assets | 0.0119*** | 0.0111*** | 0.2415*** |
| | [0.0034] | [0.0036] | [0.0417] |
| Log total assets | 0.0068*** | 0.0068*** | 0.0406*** |
| | [0.0010] | [0.0011] | [0.0081] |
| Interest share | 0.0236*** | 0.0235*** | 0.1920*** |
| | [0.0025] | [0.0025] | [0.0234] |
| Constant | -0.0300*** | -0.0301*** | -0.3083*** |
| | [0.0055] | [0.0055] | [0.0492] |
| Observations (Adj $R^2$) | 714 (0.20) | 714 (0.20) | 714 (0.16) |
| **$DIV_{FUND}$** | **-0.0258*** | **-0.0243*** | **-0.1600** |
| | [0.0169] | [0.0151] | [0.1396] |
| **$(DIV_{FUND})^2$** | **0.0592** | **0.0385** | **0.3339** |
| | [0.0432] | [0.0390] | [0.3672] |
| Bank concentration | 0.0111*** | 0.0135*** | 0.0966*** |
| | [0.0037] | [0.0034] | [0.0293] |
| Bank freedom | -0.0000* | -0.0000*** | 0.0001 |
| | [0.0000] | [0.0000] | [0.0004] |
| Equity to assets | 0.0105*** | 0.0097*** | 0.2345*** |
| | [0.0034] | [0.0035] | [0.0423] |
| Log total assets | 0.0078*** | 0.0077*** | 0.0461*** |
| | [0.0010] | [0.0011] | [0.0079] |
| Interest share | 0.0242*** | 0.0244*** | 0.1929*** |
| | [0.0026] | [0.0026] | [0.0240] |
| Constant | -0.0371*** | -0.0348*** | -0.3314*** |
| | [0.0048] | [0.0046] | [0.0429] |
| Observations (Adj $R^2$) | 714 (0.19) | 714 (0.20) | 714 (0.16) |
| **$DIV_{BAL}$** | **-0.0418*** | **-0.0391*** | **-0.0941** |
| | [0.0149] | [0.0143] | [0.1295] |
| **$(DIV_{BAL})^2$** | **0.1066*** | **0.1100*** | **0.1688** |
| | [0.0352] | [0.0338] | [0.3024] |
| Bank concentration | 0.0129*** | 0.0155*** | 0.0999*** |
| | [0.0037] | [0.0034] | [0.0296] |
| Bank freedom | -0.0001** | -0.0001*** | 0.0001 |
| | [0.0000] | [0.0000] | [0.0004] |
| Equity to assets | 0.0116*** | 0.0112*** | 0.2331*** |
| | [0.0035] | [0.0035] | [0.0413] |
| Log total assets | 0.0073*** | 0.0072*** | 0.0429*** |
| | [0.0010] | [0.0011] | [0.0079] |
| Interest share | 0.0219*** | 0.0221*** | 0.1865*** |
| | [0.0025] | [0.0025] | [0.0233] |
| Constant | -0.0346*** | -0.0327*** | -0.3225*** |
| | [0.0049] | [0.0047] | [0.0441] |
| Observations (Adj $R^2$) | 714 (0.20) | 714 (0.21) | 714 (0.16) |

Notes: Regressions used Y = [$RAR_{ROA}$, Sharpe ratio, Z-score] as dependant variables. Three equations are estimated:
(1) $Y_i = \alpha + \beta DIV_{ASS_i} + \delta DIV^2_{ASS_i} + \gamma X_i + \varepsilon_i$ ;   (2) $Y_i = \alpha + \beta DIV_{FUND_i} + \delta DIV^2_{FUND_i} + \gamma X_i + \varepsilon_i$.
(3) $Y_i = \alpha + \beta DIV_{BAL_i} + \delta DIV^2_{BAL_i} + \gamma X_i + \varepsilon_i$. Where $DIV_{ASS}$, $DIV_{FUND}$, and $DIV_{BAL}$ are respectively the assets, the funding and the balance sheet diversification's variables; X is the vector of control variables; and ε is the error terms. Dummy variables for country, years and bank type are included in all regressions but not reported. ***, **, * indicate significance at 99%, 95% and 90% level of significance respectively.

Consistent with our second hypothesis, findings support Winton (1999)'s arguments that diversification across loan sectors helps a banks return most when loans have moderate exposure to sector downturns; when loans have low downside risk, diversification have little benefit; when loans have sufficiently high downside risk, diversification may actually reduce its return.

### *6.3. The effect of revenue diversification with controlling key banking aspects*

We investigate empirically our third theoretical hypothesis **H3**: *The diversification performance's effect is inherently different with activities and across banks. There are some situations where financial institutions gains greatly from diversification. But, this depends on diverse bank specific characteristics, as well as regulatory measures*. We examine this issue by considering three banking aspects: the credit institution type, the bank specific characteristics and the regulatory restrictions. The rationale behind testing for key banking aspects is that different banks all have differing functions, restrictions and ownership structure. As a consequence, they adopt distinct approaches to diversification to achieve their strategic objectives.

### *6.3.1. Diversification revenue and the bank type*

First, we examine if bank type impacts upon diversification benefits. We retain the methodology employed in earlier regressions and additionally control for bank type by introducing dummy variables. Table14 reports estimates of the diversification interacted with dummy variables of commercial banks (column1), investment banks (column2), non-banks credit institutions (column3) and other-banks (column4).

Results show that the coefficient of the diversification interaction variables vary significantly with the bank type, contrarily to the coefficient of the non-interest income share which is found to be generally negative in all specifications. The diversification effect appears positive and quantitatively large for other-banks category, comparatively less significant for commercial banks, and insignificant for the investment banks and the non-banks credit institutions. Interestingly, the other-banks category includes highly specialized ones such as saving banks, cooperative banks, real estate and mortgage banks, medium and long term credit banks, so some activities diversification seems specifically benefic to enhance return and to lower the concentration risk. This result is aligned with the findings of Brunner et al. (2004) who report that specialized bank institutions as cooperative banks tend to reap higher returns from diversification than other banks. Diversification provides also some gains for commercial banks as measured by the risk-adjusted ROA and the Z-score. This may be due to distinct competitive advantage that those banks hold from their access to private information so as to enhance decision making.

Table14. Revenue diversification, bank type and risk-adjusted performance

| Regressors | Risk-adjusted return | | | | | | | | Z-score | | | |
|---|---|---|---|---|---|---|---|---|---|---|---|---|
| | RAR$_{ROA}$ | | | | Sharpe ratio | | | | | | | |
| **DIV $_{REV}$*Dum $_{Commer}$** | 0.0130* | | | | 0.0125 | | | | 0.2089*** | | | |
| | [0.0078] | | | | [0.0080] | | | | [0.0725] | | | |
| **DIV $_{REV}$*Dum $_{Invest}$** | | 0.0001 | | | | -0.0012 | | | | 0.0066 | | |
| | | [0.0062] | | | | [0.0065] | | | | [0.0550] | | |
| **DIV $_{REV}$*Dum $_{Non-banks}$** | | | -0.0042 | | | | 0.0001 | | | | 0.0162 | |
| | | | [0.0106] | | | | [0.0126] | | | | [0.0681] | |
| **DIV $_{REV}$*Dum $_{Other}$** | | | | 0.0537** | | | | 0.0566** | | | | 0.5791*** |
| | | | | [0.0257] | | | | [0.0246] | | | | [0.2203] |
| **SH $_{NON}$*Dum $_{Commer}$** | -0.0277*** | | | | -0.0271*** | | | | -0.2453*** | | | |
| | [0.0048] | | | | [0.0047] | | | | [0.0410] | | | |
| **SH $_{NON}$*Dum $_{Invest}$** | | -0.0077** | | | | -0.0073** | | | | -0.0814*** | | |
| | | [0.0033] | | | | [0.0035] | | | | [0.0290] | | |
| **SH $_{NON}$*Dum $_{Non-banks}$** | | | -0.0081 | | | | -0.0100 | | | | -0.1520*** | |
| | | | [0.0058] | | | | [0.0068] | | | | [0.0373] | |
| **SH $_{NON}$*Dum $_{Other}$** | | | | -0.0462** | | | | -0.0510*** | | | | -0.3657** |
| | | | | [0.0234] | | | | [0.0211] | | | | [0.1629] |
| Bank concentration | 0.0116*** | 0.0092*** | 0.0102*** | 0.0100*** | 0.0141*** | 0.0117*** | 0.0127*** | 0.0125*** | 0.0974*** | 0.0830*** | 0.0941*** | 0.0895*** |
| | [0.0038] | [0.0038] | [0.0037] | [0.0037] | [0.0033] | [0.0035] | [0.0034] | [0.0034] | [0.0306] | [0.0302] | [0.0297] | [0.0298] |
| Bank freedom | -0.0001*** | -0.0001** | -0.0001** | -0.0001** | -0.0001*** | -0.0001*** | -0.0001*** | -0.0001*** | -0.0001 | 0.0001 | -0.0001 | 0.0000 |
| | [0.0000] | [0.0000] | [0.0000] | [0.0000] | [0.0000] | [0.0000] | [0.0000] | [0.0000] | [0.0004] | [0.0004] | [0.0004] | [0.0004] |
| Equity to assets | 0.0084** | 0.0079** | 0.0077** | 0.0076** | 0.0074** | 0.0070** | 0.0067* | 0.0065* | 0.2184*** | 0.2102*** | 0.2106*** | 0.2141*** |
| | [0.0036] | [0.0035] | [0.0035] | [0.0035] | [0.0037] | [0.0037] | [0.0036] | [0.0036] | [0.0429] | [0.0439] | [0.0421] | [0.0438] |
| Log total assets | 0.0071*** | 0.0070*** | 0.0072*** | 0.0072*** | 0.0069*** | 0.0068*** | 0.0071*** | 0.0071*** | 0.0395*** | 0.0407*** | 0.0437*** | 0.0433*** |
| | [0.0010] | [0.0010] | [0.0010] | [0.0010] | [0.0011] | [0.0011] | [0.0011] | [0.0011] | [0.0081] | [0.0082] | [0.0081] | [0.0081] |
| Constant | -0.0120** | -0.0183*** | -0.0195*** | -0.0177*** | -0.0095** | -0.0157*** | -0.0167*** | -0.0152*** | -0.1555*** | -0.1794*** | -0.1945*** | -0.1696*** |
| | [0.0050] | [0.0045] | [0.0046] | [0.0045] | [0.0049] | [0.0044] | [0.0044] | [0.0045] | [0.0456] | [0.0405] | [0.0415] | [0.0404] |
| Observations (Adj R$^2$) | 714 (0.17) | 714 (0.13) | 714 (0.13) | 714 (0.13) | 714 (0.17) | 714 (0.14) | 714 (0.14) | 714 (0.14) | 714 (0.13) | 714 (0.11) | 714 (0.11) | 714 (0.11) |

Notes: Regressions used Y = [RAR$_{ROA}$, Sharpe ratio, Z-score] as dependant variables. Four equations are estimated:

(1) $Y_i = \alpha + \beta DIV_{REV_i} * Dum_{Commer_i} + \delta SH_{NON_i} * Dum_{Commer_i} + \gamma X_i + \varepsilon_i$ ;      (2) $Y_i = \alpha + \beta DIV_{REV_i} * Dum_{Invest_i} + \delta SH_{NON_i} * Dum_{Invest_i} + \gamma X_i + \varepsilon_i$  ;

(3) $Y_i = \alpha + \beta DIV_{REV_i} * Dum_{Non-banks_i} + \delta SH_{NON_i} * Dum_{Non-banks_i} + \gamma X_i + \varepsilon_i$ ;    (4) $Y_i = \alpha + \beta DIV_{REV_i} * Dum_{Other_i} + \delta SH_{NON_i} * Dum_{Other_i} + \gamma X_i + \varepsilon_i$

Where DIV$_{REV}$*Dum$_{Commer}$, DIV$_{REV}$*Dum$_{Invest}$, DIV$_{REV}$*Dum$_{Non-banks}$, and DIV$_{REV}$*Dum$_{Other}$ are respectively the diversification interaction terms with commercial banks, investment banks, non-banks, and other-banks dummy variables; SH$_{NON}$*Dum$_{Commer}$, SH$_{NON}$*Dum$_{Invest}$, SH$_{NON}$*Dum$_{Non-banks}$, and SH$_{NON}$*Dum$_{Other}$ are the share of non-interest income interacted with different bank type dummy variables; X is the vector of control variables; and ε is the error terms. The category of commercial banks includes commercial banks, and banks holding and holding companies; the category of investment banks includes investment banks and securities houses; the category of non-banking credit institutions includes non-bank credit institutions, specialized government credit institutions, and micro-financing institutions; and the category of other-banks include saving banks, cooperative banks, real estate and mortgage banks, Islamic banks, medium and long term credit banks, and multilateral government banks. Dummy variables for country, years and bank type are included in all regressions but not reported. ***, **, * indicate significance at 99%, 95% and 90% level of significance respectively.

Conforming to our third hypothesis, the empirical diversification effect is seen to be no homogeneous across bank types, mainly because banks of different types have more or less complex organization why they differ materially in both the non-interest income share and the degree of diversification. Most obviously, a bank's type entails a specific charter that outlines allowed and disallowed bank activities. Non-banking credit institutions, for instance may not be allowed to engage in investment banking activities, which limits their potential to generate non-interest income; while investment banks are naturally well diversified toward non-traditional activities, so more diversification don't exert a significant impact. Additional regressions on the different diversification aspects –assets, funding and balance sheet diversification– interacted with bank types are also conducted but not reported because providing almost similar quantitative results by bank type. These regressions provide further evidence that certain bank categories, particularly commercial banks and other-banks benefit from assets, funding and balance sheet diversification.

### 6.3.2. Diversification revenue and the bank specific characteristics

Second, we examine how the impact of diversification may vary with banks characteristics. Table15 reports estimates of regressions with the diversification revenue interacted with bank size (column1), growth (column2), profitability (column3), capitalization (column4), and efficiency (column5). Only regressions on the Sharpe ratio and the Z-score are reported, because results on risk-adjusted ROA and Sharpe ratio are almost quantitatively similar.

Results show that the diversification interaction terms enter all regressions positively, inversely to a strong negative coefficient found when the diversification variable is introduced separately. With bank specific characteristics interaction, it appears that banks tend to reap gain from revenue diversification, but this gain differs notably with banks variation in performance. Not surprisingly, efficient banks have large benefits from revenue diversification, mainly because they are able to perform new activities more efficiently. Well-capitalized banks benefit also from the risk preventing effect from diversification. This finding is consistent with the idea that high capital banks protect better their charter value. Large banks are also likely to benefit from income diversification and to improve risk-adjusted performance. One interpretation is that larger banks, which have been involved in banking activities for a longer period of time, have had time to reach the optimal level of diversification. Moreover, they are more likely to have implemented the business practices and advanced technology needed to be successful for some activities. Hunter and Timme (1986) found that larger banks are better equipped to use new technology and to exploit the resulting cost savings and/or efficiency gains. Other banks characteristics such as growth or profitability are less relevant in how to exploit the diversification potential.

Table15. Revenue diversification, bank specific characteristics and risk-adjusted performance

| Regressors | Risk-adjusted return | | | | | Z-score | | | | |
|---|---|---|---|---|---|---|---|---|---|---|
| | Sharpe ratio | | | | | | | | | |
| **DIV $_{REV}$** | **-0.0712**\*\*\* | **-0.0157**\*\*\* | **-0.0171**\*\*\* | **-0.0196**\*\*\* | **-0.0229**\*\*\* | **-0.3481**\*\*\* | **-0.0234** | **-0.0182** | **-0.1814**\*\*\* | **-0.0576** |
| | [0.0086] | [0.0056] | [0.0055] | [0.0055] | [0.0060] | [0.0790] | [0.0555] | [0.0560] | [0.0555] | [0.0567] |
| **DIV $_{REV}$* Size** | **0.0201**\*\*\* | | | | | **0.1176**\*\*\* | | | | |
| | [0.0030] | | | | | [0.0206] | | | | |
| **DIV $_{REV}$* Growth** | | **0.0024**\* | | | | | **0.0092** | | | |
| | | [0.0013] | | | | | [0.0147] | | | |
| **DIV $_{REV}$* Profitability** | | | **0.0005**\*\*\* | | | | | **-0.0006** | | |
| | | | [0.0001] | | | | | [0.0021] | | |
| **DIV $_{REV}$* Capital** | | | | **0.0143** | | | | | **0.6119**\*\*\* | |
| | | | | [0.0111] | | | | | [0.1339] | |
| **DIV $_{REV}$* Efficiency** | | | | | **0.4550**\*\*\* | | | | | **2.1256**\*\* |
| | | | | | [0.1368] | | | | | [1.0184] |
| Bank concentration | 0.0145\*\*\* | 0.0141\*\*\* | 0.0150\*\*\* | 0.0133\*\*\* | 0.0133\*\*\* | 0.1000\*\*\* | 0.0982\*\*\* | 0.0955\*\*\* | 0.0885\*\*\* | 0.0949\*\*\* |
| | [0.0033] | [0.0033] | [0.0033] | [0.0033] | [0.0033] | [0.0301] | [0.0297] | [0.0296] | [0.0296] | [0.0294] |
| Bank freedom | -0.0001\*\*\* | -0.0001\*\*\* | -0.0001\*\*\* | -0.0001\*\*\* | -0.0001\*\*\* | 0.0001 | 0.0001 | 0.0001 | 0.0002 | 0.0001 |
| | [0.0000] | [0.0000] | [0.0000] | [0.0000] | [0.0000] | [0.0004] | [0.0004] | [0.0004] | [0.0004] | [0.0004] |
| Equity to assets | 0.0072\*\* | 0.0086\*\*\* | 0.0078\*\* | | 0.0064\* | 0.2224\*\*\* | 0.2334\*\*\* | 0.2345\*\*\* | | 0.2230\*\*\* |
| | [0.0034] | [0.0036] | [0.0035] | | [0.0037] | [0.0419] | [0.0427] | [0.0428] | | [0.0433] |
| Log total assets | | 0.0072\*\*\* | 0.0076\*\*\* | 0.0068\*\*\* | 0.0066\*\*\* | | 0.0435\*\*\* | 0.0427\*\*\* | 0.0407\*\*\* | 0.0406\*\*\* |
| | | [0.0011] | [0.0011] | [0.0011] | [0.0011] | | [0.0078] | [0.0079] | [0.0079] | [0.0078] |
| Interest share | 0.0192\*\*\* | 0.0197\*\*\* | 0.0180\*\*\* | 0.0187\*\*\* | 0.0180\*\*\* | 0.1815\*\*\* | 0.1844\*\*\* | 0.1865\*\*\* | 0.1626\*\*\* | 0.1763\*\*\* |
| | [0.0025] | [0.0025] | [0.0025] | [0.0025] | [0.0026] | [0.0225] | [0.0223] | [0.0230] | [0.0220] | [0.0228] |
| Constant | -0.0082\*\* | -0.0286\*\*\* | -0.0287\*\*\* | -0.0244\*\*\* | -0.0246\*\*\* | -0.2011\*\*\* | -0.3230\*\*\* | -0.3217\*\*\* | -0.2411\*\*\* | -0.3045\*\*\* |
| | [0.0040] | [0.0054] | [0.0054] | [0.0048] | [0.0055] | [0.0404] | [0.0460] | [0.0460] | [0.0417] | [0.0465] |
| Observations (Adj R$^2$) | 714 (0.21) | 714 (0.20) | 714 (0.21) | 714 (0.20) | 714 (0.21) | 714 (0.16) | 714 (0.15) | 714 (0.15) | 714 (0.15) | 714 (0.16) |

Notes: Regressions used Y = [RAR$_{ROA}$, Sharpe ratio, Z-score] as dependant variables. Five equations are estimated: (1) $Y_i = \alpha + \beta DIV_{REV_i} + \delta DIV_{REV_i} * Size_i + \gamma X_i + \varepsilon_i$ ;
(2) $Y_i = \alpha + \beta DIV_{REV_i} + \delta DIV_{REV_i} * Growth_i + \gamma X_i + \varepsilon_i$ ; (3) $Y_i = \alpha + \beta DIV_{REV_i} + \delta DIV_{REV_i} * Profitability_i + \gamma X_i + \varepsilon_i$ ;
(4) $Y_i = \alpha + \beta DIV_{REV_i} + \delta DIV_{REV_i} * Capital_i + \gamma X_i + \varepsilon_i$ ; (5) $Y_i = \alpha + \beta DIV_{REV_i} + \delta DIV_{REV_i} * Efficiency_i + \gamma X_i + \varepsilon_i$.
Where DIV$_{REV}$*Size$_r$, DIV$_{REV}$*Growth, DIV$_{REV}$*Profitability, DIV$_{REV}$*Capital, and DIV$_{REV}$*Efficiency are respectively the diversification interaction terms with size, growth, profitability, capital, and efficiency; X is the vector of control variables; and ε is the error terms. Bank size, growth, profitability, capital and efficiency are controlled respectively by Log total assets, assets growth, net interest margin, equity to assets, and income to cost ratio. Dummy variables for country, years and bank type are included in all regressions but not reported. \*\*\*, \*\*, \* indicate significance at 99%, 95% and 90% level of significance respectively.

Our findings imply that very large, well-capitalized and more efficient banks have more incentives to diversify. They perform better than the other or the specialized ones; traditional forms of intermediation are less profitable for them. Extended regressions on the different diversification forms –assets, funding and balance sheet diversification– interacted with bank specific characteristics are also conducted but not reported due to similarly quantitative results. Regressions consistent with our third hypothesis, confirm that banks reap profit and stability gains from assets, funding and balance sheet diversification, particularly when they are large, highly capitalized and efficient.

*6.3.3. Diversification revenue and the regulatory restrictions*

Finally, we check how differences in bank regulation may further affect performance. We retain the methodology employed in previous regressions and additionally control for regulatory restrictions by introducing interaction terms. Table16 reports estimates of the diversification interacted with securities restrictions (column1), insurance restrictions (column2), real estate restrictions (column3), and hazard moral mitigating factors (column 4).

Results on the risk-adjusted returns indicate that if banks face more enforcement in restrictions on securities or real estate activities or in factors mitigating hazard moral as they diversify to non-traditional activities, they avoid the negative diversification effect as shown by the insignificant coefficient of the diversification interaction terms. In contrast, restrictions on insurance strengthen the poor diversification return of banks, as seen by the negative correlation between diversification interaction term and the Sharpe ratio. This suggests that banks can reap benefits from consolidating banking and insurance agency activities. According to Chang and Elyasini (2008), we find that synergies do exist between banking and insurance activities. Contrarily to the securities or real estate businesses that imply cost to develop skills and expertise to handle uncertain risk, the insurance activities require comparatively a low level of investment and they are more profitable than banking services. As Saunders and Walter (1994) have pointed out, certain insurance products, e.g. credit life insurance, mortgage insurance, and auto insurance tend to have natural synergistic links to bank lending and they can be viewed as a continuity of the bank traditional activities. The hazard moral result also interesting is conforming to the agency theory; mitigating moral hazard limit the agency problems from diversification. Regarding the Z-score regressions, results show that more restrictions on securities activities or an enforcement of factors mitigating hazard moral lead to increase bank insolvency risk, when other activities restrictions don't have a significant effect. Not surprisingly, financial consolidation across securities activities offer some opportunities for lowering the portfolio risk and the likelihood of financial firm failure, while factors mitigating hazard moral, generally by limiting the explicit deposit insurance guarantee or the government intervention as a lending in last resort, may decrease the scope of safety net.

Table16. Diversification revenue, regulatory restrictions and risk-adjusted performance

| Regressors | Risk-adjusted return | | | | | | | | Z-score | | | |
|---|---|---|---|---|---|---|---|---|---|---|---|---|
| | $RAR_{ROA}$ | | | | Sharpe ratio | | | | | | | |
| **$DIV_{REV}$** | -0.0107 | 0.0008 | -0.0196** | -0.0179* | -0.0143** | 0.0159 | -0.0165* | -0.0301*** | 0.0880 | -0.0756 | -0.0301 | 0.1065 |
| | [0.0069] | [0.0122] | [0.0084] | [0.0109] | [0.0071] | [0.0137] | [0.0089] | [0.0103] | [0.0685] | [0.1156] | [0.0810] | [0.0922] |
| **$DIV_{REV}$* Securities** | -0.0024 | | | | -0.0003 | | | | -0.0576*** | | | |
| | [0.0020] | | | | [0.0023] | | | | [0.0172] | | | |
| **$DIV_{REV}$* Insurance** | | -0.0060 | | | | -0.0115*** | | | | 0.0203 | | |
| | | [0.0041] | | | | [0.0044] | | | | [0.0342] | | |
| **$DIV_{REV}$* Real-estate** | | | 0.0016 | | | | 0.0005 | | | | 0.0035 | |
| | | | [0.0026] | | | | [0.0027] | | | | [0.0213] | |
| **$DIV_{REV}$* Hazard** | | | | -0.0070 | | | | 0.0009 | | | | -0.1242*** |
| | | | | [0.0059] | | | | [0.0056] | | | | [0.0451] |
| Bank concentration | 0.0121*** | 0.0115*** | 0.0111*** | 0.0230*** | 0.0139*** | 0.0141*** | 0.0137*** | 0.0230*** | 0.1144*** | 0.0964*** | 0.0966*** | 0.2018*** |
| | [0.0037] | [0.0037] | [0.0036] | [0.0077] | [0.0033] | [0.0033] | [0.0033] | [0.0070] | [0.0298] | [0.0291] | [0.0288] | [0.0550] |
| Bank freedom | -0.0001** | -0.0001*** | -0.0000 | 0.0000 | -0.0001*** | -0.0001*** | -0.0001*** | -0.0001 | -0.0004 | 0.0002 | 0.0001 | 0.0021** |
| | [0.0000] | [0.0000] | [0.0000] | [0.0001] | [0.0000] | [0.0000] | [0.0000] | [0.0000] | [0.0005] | [0.0004] | [0.0005] | [0.0010] |
| Equity to assets | 0.0090*** | 0.0096*** | 0.0097*** | 0.0099** | 0.0085*** | 0.0087*** | 0.0087*** | 0.0090* | 0.2213*** | 0.2333*** | 0.2338*** | 0.2140*** |
| | [0.0034] | [0.0034] | [0.0034] | [0.0046] | [0.0036] | [0.0035] | [0.0036] | [0.0049] | [0.0432] | [0.0427] | [0.0428] | [0.0585] |
| Log total assets | 0.0072*** | 0.0075*** | 0.0072*** | 0.0071*** | 0.0072*** | 0.0076*** | 0.0072*** | 0.0077*** | 0.0411*** | 0.0426*** | 0.0431*** | 0.0327*** |
| | [0.0009] | [0.0009] | [0.0009] | [0.0014] | [0.0011] | [0.0011] | [0.0011] | [0.0016] | [0.0079] | [0.0076] | [0.0076] | [0.0106] |
| Interest share | 0.0190*** | 0.0186*** | 0.0200*** | 0.0196*** | 0.0196*** | 0.0178*** | 0.0198*** | 0.0211*** | 0.1698*** | 0.1877*** | 0.1853*** | 0.1529*** |
| | [0.0025] | [0.0026] | [0.0025] | [0.0034] | [0.0026] | [0.0026] | [0.0026] | [0.0035] | [0.0227] | [0.0233] | [0.02322] | [0.0282] |
| Constant | -0.0291*** | -0.0288*** | -0.0326*** | -0.0427*** | -0.0281*** | -0.0245*** | -0.0290*** | -0.0354*** | -0.2834*** | -0.3288*** | -0.3260*** | -0.4447*** |
| | [0.0054] | [0.0054] | [0.0062] | [0.0112] | [0.0055] | [0.0054] | [0.0062] | [0.0106] | [0.0477] | [0.0478] | [0.0540] | [0.0836] |
| Observations (Adj $R^2$) | 714 (0.20) | 714 (0.20) | 714 (0.20) | 714 (0.23) | 714 (0.20) | 714 (0.20) | 714 (0.20) | 714 (0.22) | 714 (0.16) | 714 (0.15) | 714 (0.15) | 714 (0.18) |

Notes: Regressions used Y = [$RAR_{ROA}$, Sharpe ratio, Z-score] as dependant variables. Four equations are estimated:
(1) $Y_i = \alpha + \beta DIV_{REV_i} + \delta DIV_{REV_i} * Securities_i + \gamma X_i + \varepsilon_i$ ; (2) $Y_i = \alpha + \beta DIV_{REV_i} + \delta DIV_{REV_i} * Insurance_i + \gamma X_i + \varepsilon_i$ ;
(3) $Y_i = \alpha + \beta DIV_{REV_i} + \delta DIV_{REV_i} * Real-estate_i + \gamma X_i + \varepsilon_i$ ; (4) $Y_i = \alpha + \beta DIV_{REV_i} + \delta DIV_{REV_i} * Hazard_i + \gamma X_i + \varepsilon_i$.
Where $DIV_{REV}$*Securities, $DIV_{REV}$*Insurance, $DIV_{REV}$*Real-estate, and $DIV_{REV}$*Hazard are respectively the diversification interaction terms with indexes of securities, insurance and real-estate restrictions, and hazard moral mitigating; X is the vector of control variables; and ε is the error terms. Regulatory variables are from Ross Levine database (2003). Dummy variables for country, years and bank type are included in all regressions but not reported. ***, **, * indicate significance at 99%, 95% and 90% level of significance respectively.

In sum, regulatory restrictions are seen to drive different diversification effects on bank performance, further validating our third hypothesis. This differential effect is robust to the diverse forms of revenue diversification. Additional tests on assets, funding and balance sheet diversification combined with regulatory restrictions confirm our earlier findings[6]. Insurance activities may provide a beneficial area in which banks can engage, but securities and real-estate activities must be regulated. At the same time, the Z-score still adversely affected by the regulation of both securities activities and hazard moral.

## 7. Concluding remarks

The contribution of this paper is to study the implications of the recent banks involvement in emerging market economies toward non-traditional intermediation activities that generate non-interest income. While traditional banking theory recommends that the optimal organization of a bank is one where it is fully diversified, our results suggest that empirically, there seem to be diseconomies of diversification for certain banks. These diseconomies arise in the form of poor monitoring incentives that induce greater risk of default and decrease return when a bank expands into industries where it faces lack skills and expertise.

Evidence suggests that diversification gains are more than offset by the cost of increased exposure to the non-interest based activities, more specifically by the trading income volatility. More generally, the relationship between diversification and performance is found to be non linear; it's conditioned by the risk level, hence diversification increase performance only at moderate levels of risk. Moreover, this diversification performance's effect is not uniform among banks or across business lines. Our findings suggest that very large, well-capitalized and more efficient banks perform better than the other or the specialized ones; traditional forms of intermediation are less profitable for them. Also for certain categories of banks, particularly highly specialized ones, some activities diversification seems particularly benefic to improve return and to lower the concentration risk. In addition, banks can boost performance by choosing the appropriate non-banking activities for diversification. In this way, insurance activities may provide a beneficial area in which different banks can engage.

An implication of this analysis is that bank diversification into non-traditional activities should be not hazardous. Banking institutions can reap diversification benefits as long as they well-studied it and know just how much diversification would be necessary to achieve positive result by considering its specific characteristics, capabilities and the risk level, and as they choose the right niche.

---

[6] Additional tests are not reported because providing similar results.

# Appendix

*Table1. Overview of countries surveyed*

| Country | Number of banks over the sample period | Number of listed banks |
|---|---|---|
| *Latin-America* | | |
| Argentina | 85 | 5 |
| Brazil | 141 | 18 |
| Chile | 23 | 7 |
| Colombia | 29 | 2 |
| Ecuador | 24 | 0 |
| Mexico | 42 | 3 |
| Venezuela | 44 | 5 |
| *East-Asia* | | |
| Indonesia | 49 | 22 |
| Hong Kong | 72 | 12 |
| Korea | 34 | 20 |
| Malaysia | 64 | 16 |
| Philippines | 37 | 13 |
| Singapore | 33 | 9 |
| Thailand | 37 | 26 |
| Total (14) | 714 | 158 |

*Table2. Overview of bank categories*

| | Criteria of classification | Number |
|---|---|---|
| *Categories of bank size* | | |
| Very large banks | Total assets > 10 Billion USD | 133 |
| Large banks | Total assets comprised between 1 and 10 Billion | 267 |
| Small banks | TA comprised between 100 Million and 1 Billion USD | 234 |
| Very small banks | Total assets < 100 Million USD | 81 |
| *Categories of bank type* | | |
| Commercial banks | Commercial banks; Bank holdings and holding companies | 521 |
| Investment banks | Investment banks and securities houses | 86 |
| Non-banking credit institutions | Non-bank credit inst. ; Specialized government credit inst.; Micro-financing inst. | 75 |
| Other banks | Saving banks; Cooperative banks; Real estate and mortgage banks; Islamic banks; Medium and long term credit banks; Multilateral government banks | 33 |

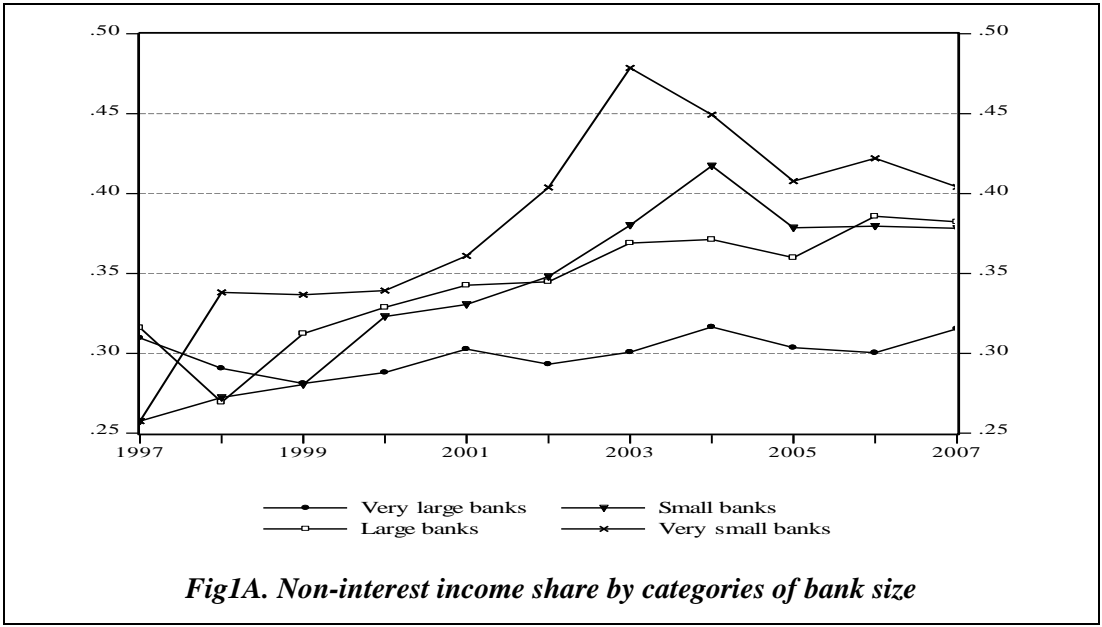
*Fig1A. Non-interest income share by categories of bank size*

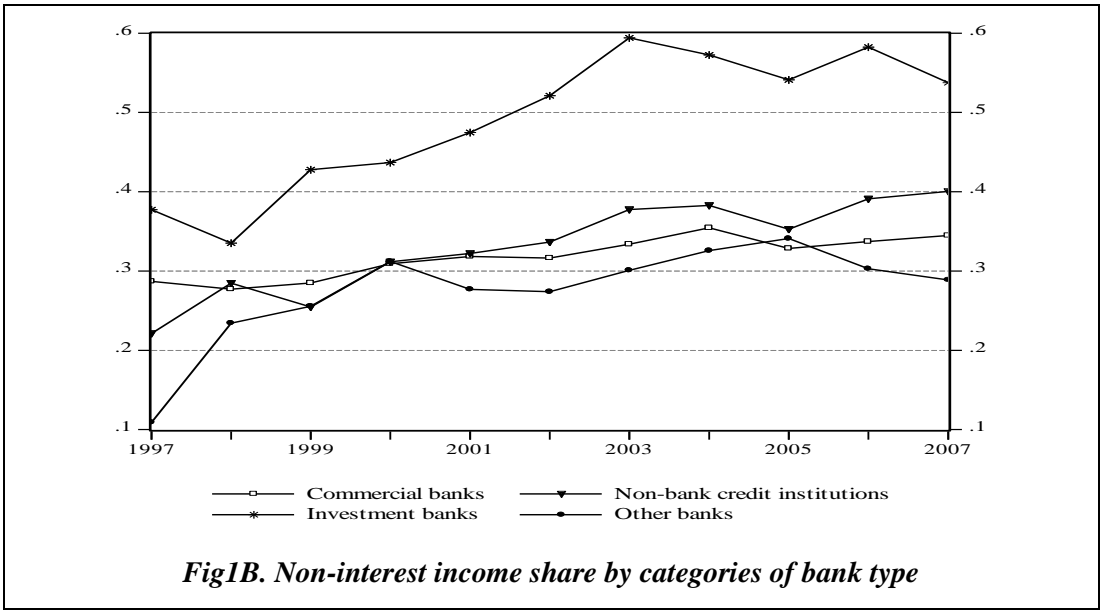
*Fig1B. Non-interest income share by categories of bank type*

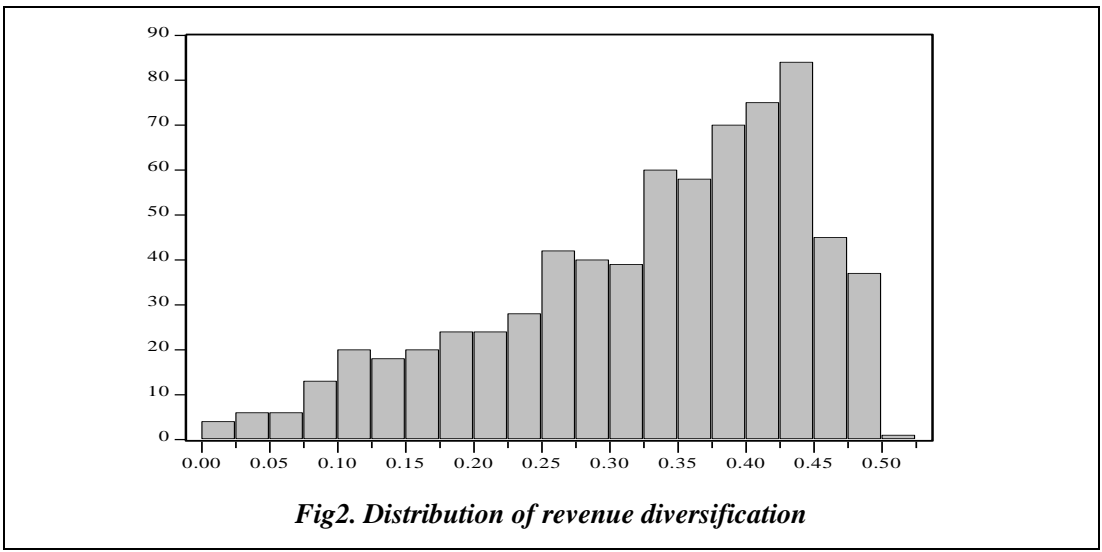
*Fig2. Distribution of revenue diversification*